\documentclass[sn-mathphys,Numbered]{sn-jnl}

\usepackage{graphicx}%
\usepackage{multirow}%
\usepackage{amsmath,amssymb,amsfonts}%
\usepackage{amsthm}%
\usepackage{mathrsfs}%
\usepackage[title]{appendix}%
\usepackage{xcolor}%
\usepackage{textcomp}%
\usepackage{manyfoot}%
\usepackage{booktabs}%
\usepackage{algorithm}%
\usepackage{algorithmicx}%
\usepackage{algpseudocode}%
\usepackage{listings}%
\usepackage{tikz}
\usepackage{cleveref}

\let\svthefootnote\thefootnote
\newcommand\freefootnote[1]{%
  \let\thefootnote\relax%
  \footnotetext{#1}%
  \let\thefootnote\svthefootnote%
}

\newcommand{\cir}[1]{\tikz[baseline]{%
    \node[anchor=base, draw, circle, inner sep=0, minimum width=1.2em]{#1};}}
    
\theoremstyle{thmstyleone}%
\theoremstyle{thmstyletwo}%

\theoremstyle{thmstylethree}%

\newcommand{\etal}{\textit{et al}.~}

\begin{document}

\title{Accurate Fine-Grained Segmentation of Human Anatomy in Radiographs via Volumetric Pseudo-Labeling
}


\author*[1,2,3]{\fnm{Constantin} \sur{Seibold}}\email{constantin.seibold@kit.edu}

\author[1]{\fnm{Alexander} \sur{Jaus}}\email{alexander.jauss@kit.edu}

\author[4]{\fnm{Matthias A.} \sur{Fink}}\email{matthias.fink@uni-heidelberg.de}

\author[2]{\fnm{Moon} \sur{Kim}}\email{moon.kim@uk-essen.de}

\author[1]{\fnm{Simon} \sur{Rei{\ss}}}\email{simon.reiss@kit.edu}

\author[3]{\fnm{Ken} \sur{Herrmann}}\email{Ken.Herrmann@uk-essen.de}

\author[2]{\fnm{Jens} \sur{Kleesiek}\footnotemark[2] $\,$}\email{jens.kleesiek@uk-essen.de}

\author[1]{\fnm{Rainer} \sur{Stiefelhagen}\footnotemark[2] $\,$}\email{rainer.stiefelhagen@kit.edu}

\affil*[1]{\orgdiv{Institute for Anthropomatics and Robotics}, \orgname{Karlsruhe Institute of Technology}, \city{Karlsruhe}, \country{Germany}}
\affil[2]{\orgdiv{Institute for Artificial Intelligence in Medicine}, \orgname{University Hospital Essen}, \city{Essen},  \country{Germany}}
\affil[3]{\orgdiv{Clinic for Nuclear Medicine}, \orgname{University Hospital Essen}, \city{Essen},  \country{Germany}}
\affil[4]{\orgdiv{Clinic for Diagnostic and Interventional Radiology}, \orgname{University Hospital Heidelberg}, \orgaddress{\city{Heidelberg}, \country{Germany}}}

\abstract{
\textbf{Purpose:} 
Interpreting chest radiographs (CXR) remains challenging due to the ambiguity of overlapping structures such as the lungs, heart, and bones. To address this issue, we propose a novel method for extracting fine-grained anatomical structures in CXR using pseudo-labeling of three-dimensional computed tomography (CT) scans. 

\textbf{Methods:}
We created a large-scale dataset of 10,021 thoracic CTs with 157 labels and applied an ensemble of 3D anatomy segmentation models to extract anatomical pseudo-labels. These labels were projected onto a two-dimensional plane, similar to the CXR, allowing the training of detailed semantic segmentation models for CXR without any manual annotation effort.

\textbf{Results:}
Our resulting segmentation models demonstrated remarkable performance on CXR, with a high average model-annotator agreement between two radiologists with mIoU scores of 0.93 and 0.85 for frontal and lateral anatomy, while inter-annotator agreement remained at 0.95 and 0.83 mIoU. Our anatomical segmentations allowed for the accurate extraction of relevant explainable medical features such as the cardio-thoracic-ratio.

\textbf{Conclusion:}
Our method of volumetric pseudo-labeling paired with CT projection offers a promising approach for detailed anatomical segmentation of CXR with a high agreement with human annotators. This technique may have important clinical implications, particularly in the analysis of various thoracic pathologies.

}



\maketitle

\freefootnote{\dag \phantom{,}denotes equal contribution.}

Chest radiographs (CXR) are one of the most common diagnostic imaging methods for patients with respiratory or cardiovascular conditions, with more than 130 million studies performed annually in Germany alone~\cite{bfs}. By using ionizing radiation to penetrate the body, CXR provide a visual representation of the organs, tissues and cavities and their current state. The interpretation of CXR these images is challenging, since it requires a thorough understanding of human anatomy due to the presence of overlapping structures that can obscure potential pathological changes and other abnormalities. Despite these challenges, CXR remain a standard diagnostic procedure and its quantitative analysis can be time consuming.
With the increasing demand in imaging procedures and the massive workload that comes along with it, this can lead to avoidable errors due to rushed examination~\cite{brady2017error, SOKOLOVSKAYA2015683} or burnout due to the overly straining of doctors~\cite{burnout1,burnout2,grosser2014burnout}.

Recent advances in Computer Vision, such as convolutional neural networks (CNN) or vision transformers (ViT),  have the potential to reduce the workload of radiologists in both image analysis and reporting~\cite{rezazade2021applications, KAPOOR20201363}. Several recent computer-aided diagnostic approaches automatically predict the presence or absence of relevant findings~\cite{seibold2020self}. They use study-level labels parsed from medical reports as training material. While this enables the classification of predefined abnormalities, this implicit training hinders their interpretability and usability as they are unable to concretely define their location or provide quantitative information. 
Dense pixel-wise predictors can circumvent these problems as they lead to concrete delineations of relevant structures. 
However, such models require box or mask annotations which are difficult to acquire not only because of the effort pixel-wise annotations entail, but also due to CXR presenting a summation image of overlapping structures such as (i.e. the ribs, lungs, and heart). This can make it challenging for radiologists to reach a proper consensus~\cite{seibold2022reference}. This difficulty results in immense time requirements for the accurate annotation of CXR, making the generation of manually annotated datasets sufficiently large enough to train deep learning models nigh impossible.

We propose a method to collect large amounts of annotations for CXR without manual labeling and demonstrate how these can generate networks capable of identifying thoracic anatomy in a fine-grained manner. Since manual annotation is not feasible, we are inspired from three related facts: First, CT scans are aggregated multi-view 2D radiographs and thus share a very similar origin signal~\cite{herman2009fundamentals}. Second, CTs offer immense advantages over CXRs in identifying structures due to their volumetric nature\cite{chapman2016clinical}. Finally, the consistent body structure of a patient across modalities implies that semantic can be translates from CT to CXR. 
Looking at other imaging domains, we find most datasets focus on a specific subset relevant to their specialty, e.g. segmentations of thoracic organs at risk~\cite{lambert2020segthor} or the spine~\cite{sekuboyina2021verse} in CT volumes. These datasets, often containing less than 100 patients, allow the development of robust networks that can be applied to new data~\cite{isensee2021nnu}. We hypothesize that by combining different volumetric annotations, we can obtain a holistic view on the human body that can be transferred onto a large CT corpus.

To facilitate network training for CXR segmentation the annotations and their associated images have to be projected onto a two-dimensional (2D) plane. A simple concept is the Ray-summation~\cite{suzuki2017image}, which allow one to quickly gain insight into a patient without requiring a thorough look through the CT. Matsubara~\etal alter this approach via a separate bone window and non-linearity~\cite{matsubara2019generation}. Other concurrent work approaches CT-Projection also via generative models~\cite{gao2023synthetic}. We adapt Matsubara~\etal's method via histogram equalizations to generate more realistic projections without falling into the potential pitfalls of deep unpaired generative models such as mode collapse or the prediction of unfitting patterns~\cite{zhu2017unpaired,bau2019seeing}. 

Following this process, we build the Projected Anatomy for X-Ray~\cite{Seibold_2022_BMVC} (PAXRay++) dataset containing over two million annotated instances across 15 thousand images for 157 anatomical classes for both lateral and frontal thoracic views. PAXRay++ allows us to train data-intensive 2D chest X-Ray anatomy segmentation (CXAS) models that accurately delineate the fine-grained thoracic anatomy. 

We evaluate these models in several ways. We use a hand-selected set of projected X-Rays as a validation set to select hyperparameters such as model architecture. The best model on the validation set is used for comparison with two radiologists. In a direct comparison to human annotations of 157 anatomical classes in 30 frontal and lateral images, we show a high model-annotator agreement (MAA) with the two radiologists, with average Intersection-over-Union (IoU) scores of 0.93 and 0.96 for the frontal view (anterior-posterior [AP]/ posterior-anterior [PA]) with an inter-annotator agreement (IAA) of 0.93. For the lateral view, the MAA comes to 0.83 and 0.87 with an IAA of 0.83.
We also show correlations of automatically extracted biomarkers with radiological findings such as scoliosis ($N=2054/231007$) or cardiomegaly ($N=41633/231007$) on the MIMIC-CXR dataset~\cite{johnson2019mimic} with a t-statistic of 35.30 ($\textit{p-value} < 0.0001$) and 139.71 ($\textit{p-value} < 0.0001$) respectively.

We make our models available on \href{cxas.ikim.nrw}{this project page}.

\section*{Results}
\begin{figure}
    \centering
    \includegraphics[width=\linewidth]{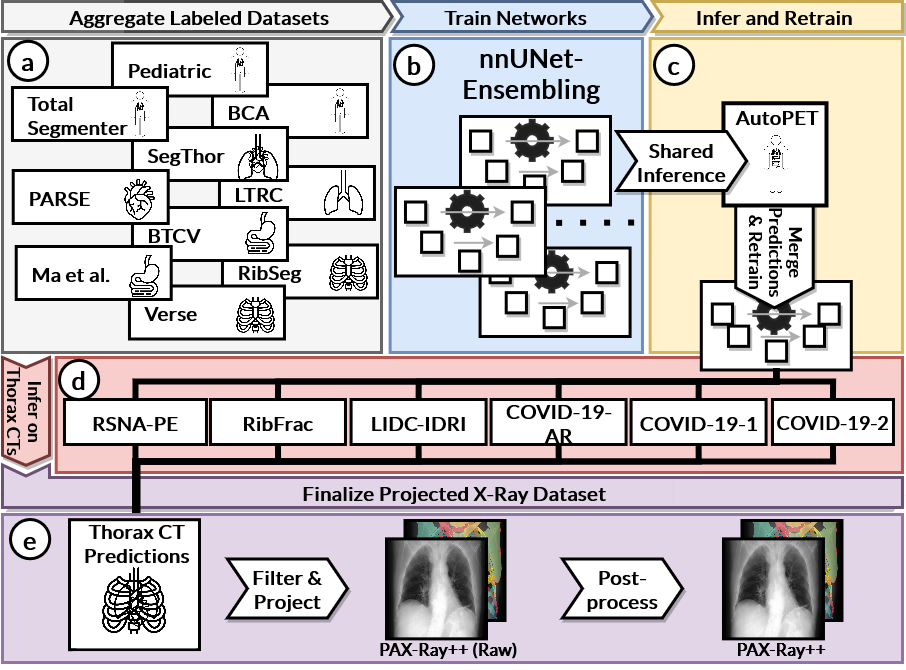}
    \caption{A flowchart describing the PAX-Ray++ dataset generation process. (a)  We collect publicly available datasets containing different anatomical structures. (b)  We train an ensemble of nnUNets of each expert domain and infer them on a shared dataset. (c)  We merge the 3D predictions, apply anatomical priors and retrain. (d)  We infer the final nnUNet on 10K chest CTs. (e)  We apply a CT and label projection to generate a chest X-Ray dataset which we apply anatomical-prior-based postprocessing to collect the final dataset.}
    \label{fig:workflow}
\end{figure}

\subsection*{Projected Chest X-Ray Anatomy Dataset Generation \\ via Volumetric Pseudo-Labeling}
Gathering vast amounts of pixel-wise annotated image data is tedious and difficult, especially in the medical domain, where we require expert annotators with a certain amount of fundamental knowledge.
To combat this, we propose a general workflow for an automated large-scale dataset generation scheme, which we visualize in Fig.~\ref{fig:workflow}. First~$\cir{a}$, we collect several datasets~\cite{jordan2022pediatric, wasserthal2022totalsegmentator, lambert2020segthor, parse, malabdomen, ma2021abdomenct, yang2021ribseg, sekuboyina2021verse} which contain anatomical structures visible in a chest X-Ray which come down to individual \emph{bones} such as ribs and vertebrae, \emph{abdominal} organs such as the liver, \emph{cardiovascular} structures such as the aorta or the heart, and \emph{respiratory} structures like lung lobes or the trachea. Subsequently~$\cir{b}$, we train an ensemble of nnUNets~\cite{isensee2021nnu} on each CT dataset. This architecture has proven itself as a reliable out-of-the-box segmentation model able to achieve strong performances with even small amounts of data, and the lack of manual hyperparameter setting allows us easily apply it for different datasets.~$\cir{c}$ After training the nnUNet ensembles on the individual datasets, we infer these expert models on a  subset of 560 whole-body volumes from AutoPET~\cite{gatidis2022whole}, a large CT dataset for tumor segmentation covering different body regions. We chose whole-body CTs because they do not impair the predictive ability of models for lower body regions such as the abdomen or lumbar spine. We merge classes of overlapping semantics and discard unnecessary classes or those with insufficient quality by manual assessment, such as the humerus. We further apply postprocessing techniques in the form of morphological operations and persistent anatomical biases like maintaining the largest connected component for organs that should contain a single one. 
We then retrain the nnUNet on the resulting dataset. The resulting model allows us to generate masks for the complete label set with a single prediction, which is significant as inference and postprocessing for an expert model can span up to an hour per volume. Thus, we mitigate a bottleneck in our workflow by compressing this knowledge into a single model.
$\cir{d}$ We collect several large-scale chest CT datasets~\cite{colak2021rsna, ribfrac2020, armato2011lung, harmon2020artificial, kassin2021generalized} ranging in size from 176 to 7,302 volumes. The combined set for inference contains 10,021 CT scans. The final nnUNet ensemble trained on AutoPET pseudo-labels generates anatomical predictions for all provided scans. 
$\cir{e}$ Subsequently, we automatically filter these volumetric pseudo-labels based on their plausibility through checks like deviation of mask center position or volume to the class mean. We discard scans that do not follow these rules. 
We add classes like lung zones, mediastinum, or aortic arch based on general rules following expert knowledge as they provide helpful insights for downstream tasks. Afterward, we project the image and masks to a frontal and lateral view and resize them to a uniform size.
For the CT projection, we adapt Matsubara\etal~\cite{matsubara2019generation} by removing the imaging table and applying histogram equalization~\cite{pizer1987adaptive} to generate more visually pleasing results. 
The masks are then post-processed using standard morphological operations. 

Following this procedure, we generate the PAX-Ray++ dataset with a total of 14,754 images, 157 densely labeled anatomy classes, and more than two million instance mask annotations, the currently largest dataset for CXR segmentation to date. We discuss the details of our dataset in the following.

We present details of the used datasets and the performance of the nnUNet on validation sets in Tab.~\ref{tab:ch5_datasets_overview}.

\begin{figure}[!]
    \centering
    \includegraphics[width=\linewidth]{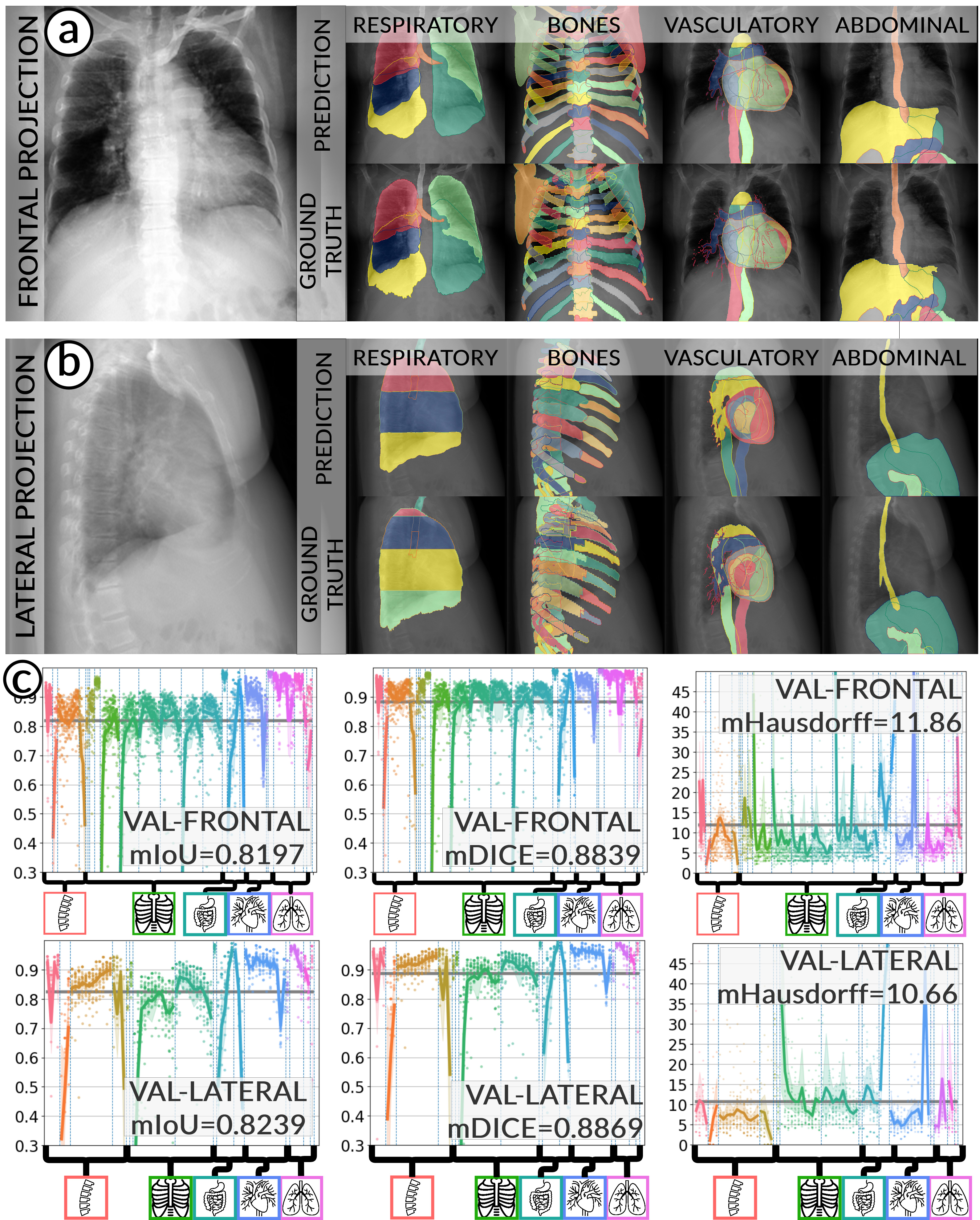}
    \caption{a) We show qualitative results for our CXAS model on frontal projections. We highlight masks for the respiratory system, bones, vasculatory system and abdomen. a) We show qualitative results for our CXAS model on lateral projections. We highlight masks for the respiratory system, bones, vasculatory system and abdomen. c) We show the performance of our  CXAS model in terms of mIoU, mDICE and mHausdorff distance for frontal (top row) and lateral (bottom row) images. }
    \label{fig:validation}
\end{figure}

\subsection*{Evaluation of CXAS Performance}
For all our experiments shown for chest X-Ray anatomy segmentation (CXAS), we use a UNet~\cite{ronneberger2015u} with a pre-trained ResNet-50 backbone~\cite{he2016deep}. We train the model on the PAX-Ray++ dataset and show details of the training choices in the Methods.

\subsubsection*{Comparison of CXAS with CT Projections}

We randomly select a validation set of 5 CTs from each source CT dataset of different patients. We manually inspect the associated frontal and lateral images to be visual correctness. The resulting validation set consists of a total of 60 images with 157 labels. We use the validation set for the development of the segmentation model and setting of the hyperparameters.
We evaluate our model via the mean Intersection-over-Union, DICE, and Hausdorff distance and show the performance of the final model.

We show qualitative results for frontal projections in Fig.~\ref{fig:validation}$\cir{a}$. We show a subset of classes belonging to the supercategories lungs, vascular systems, bones, and abdomen/digestive system. The predictions show minor deviations at the boundaries of the individual classes of the respiratory and vasculature system, while some inaccuracies become visible in the abdominal area. 
The qualitative results for the lateral projections are displayed in Fig.~\ref{fig:validation}$\cir{b}$. Akin to the frontal view, the predictions show smoother borders but align with the ground truth. 
Apart from this, the segmentations provide matching insights on the thoracic anatomy with a slight deviation from the ground truth for both frontal and lateral views. 

Fig.~\ref{fig:validation}$\cir{c}$ shows quantitative segmentation results for frontal (top row) and lateral (bottom row) views. We display the class performances in the form of IoU (left), DICE (center), and Hausdorff distance (right) for each sample as a scatter plot with the mean performance for the classes shown by a line plot. 
Generally, we see performances for standard spinal classes, such as the thoracic vertebrae with average IoU-scores above 80\%, while the average performance of rare vertebrae of the dataset belonging to the cervical and lumbar spine can drop down to 40\%. In the frontal view, there exists more variance in thoracic vertebrae segmentation performance compared to the lateral view. 
Bone structures such as the sternum, clavicles, and scapula achieve IoUs in the mean from 85\% to 95\%. 
For ribs, we can see a noticeable performance drop for the anterior parts of the lower ribs independent of the side. The lower anterior ribs typically do not contain a large area, making them difficult to segment. This behavior is mirrored in the lateral view across the metrics. Abdominal classes can vary in segmentation quality as they occur in a nearly homogenous region. For example, while the liver or stomach are typically well-segmented, the duodenum and kidneys are more complex. Heart and Lung related classes show near-perfect segmentations with scores above 90\% IoU. Breast tissue segmentation in comparison only achieves a mean of 70\% mIoU. 
It can be noted that classes in the lateral view tend to have slightly better scores than their frontal counterparts.

\subsubsection*{Comparison of CXAS to Human Anatomy Annotations on Real World Chest X-Rays}
\begin{figure}[!]
\includegraphics[width=\linewidth]{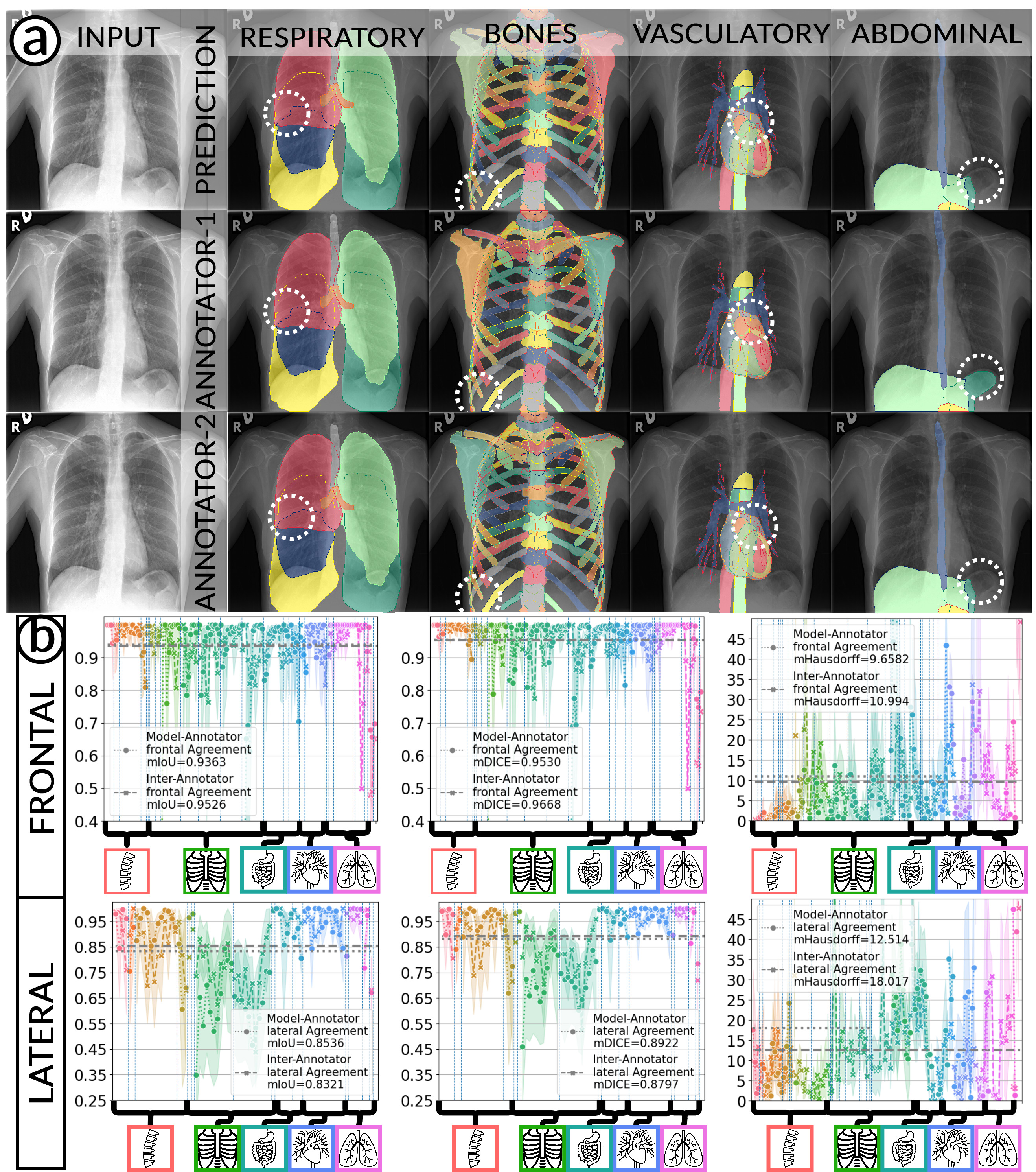}
    \caption{a) We show qualitative results for our CXAS model on frontal CXR as well as two expert manual annotations. We highlight masks for the respiratory system, bones, vasculatory system and abdomen. b) We show the performance of our  CXAS model in terms of mIoU, mDICE and mHausdorff distance for frontal (top row) and lateral (bottom row) images. } 
    \label{fig:frontal_qual_real}
\end{figure}

To test the real-world applicability of PAX-Ray++ for developing anatomy segmentation models, we prepare a test set of 30 frontal and lateral chest X-Rays each from the PadChest dataset~\cite{bustos2019padchest} for a total fo 60 images. First, we use our best segmentation model based on the validation performance to generate preliminary anatomical segmentations. Then, we tasked two medical experts to independently examine and correct these densely annotated X-Rays using the annotation tool CVAT~\cite{cvat}. Even in this simplified annotation process, the expert denoted that a single image correction would take roughly 1.5 hours on average due to the sheer amount of considered classes. 

For the inter-annotator agreement (IAA), we calculate the three prior metrics between the masks from both annotators due to their symmetrical properties. For the model-annotator agreement (MAA), we compute the metrics between the individual annotator and the model predictions and average the results class-wise. 

We display qualitative results in Fig.~\ref{fig:frontal_qual_real}$\cir{a}$.
The annotators tend to be content with most annotations. There are edits at the extensions of the esophagus, trachea, and aorta and corrections of the lower ribs. There is little consensus for classes in the abdominal area, such as the stomach, as seen on the right of Fig.~\ref {fig:frontal_qual_real}~$\cir{a}$. In contrast, the annotators often align for bone classes.

We show the quantitative segmentation performance against medical expert annotations for all classes in Fig.~\ref{fig:frontal_qual_real}~$\cir{b}$. The dashed lines represent the IAA, whereas the dotted line displays the average MAA. As the task for the human annotators was not to annotate from scratch but to correct wrong pixel-wise predictions, we can see a high MAA for most classes. In the frontal view, the most significant disagreements exist for rare bone structures such as L3 and C4, lower ribs, the mediastinal distribution, or the breast tissue. The IAA and MAA are highly similar with mIoU of respective 95\% and 94\%. The Hausdorff distance for MAA is slightly lower than the IAA, indicating slight differences in boundary annotations while maintaining a considerable overlap with the other annotator.
In the lateral view, the concrete delineation of rib structures appears ambiguous, leading to lower MAA and IAA with a greater IAA than MAA for all metrics in this supercategory. Overall there is less agreement between the medical experts in the lateral view, leading to a better average MAA than IAA across all metrics (i.e. 85\% vs 83\% mIoU). 
While the experts propose changes to the original predictions, they are often not overlapping. 
In the lateral view, rib segmentations can become quite hard to interpret. While both annotators disagree with the rib segmentations, they do not always agree on how they should look. 
Similarly to the frontal view, tube-like structures like the esophagus are extended as they can appear fractured at times.

\subsection*{Applications of Anatomy-based Feature Extraction}

\subsubsection*{Automated Identification Cardio-Thoracic Ratio via CXAS}
\begin{figure}
    \centering
    \includegraphics[width=0.95\linewidth]{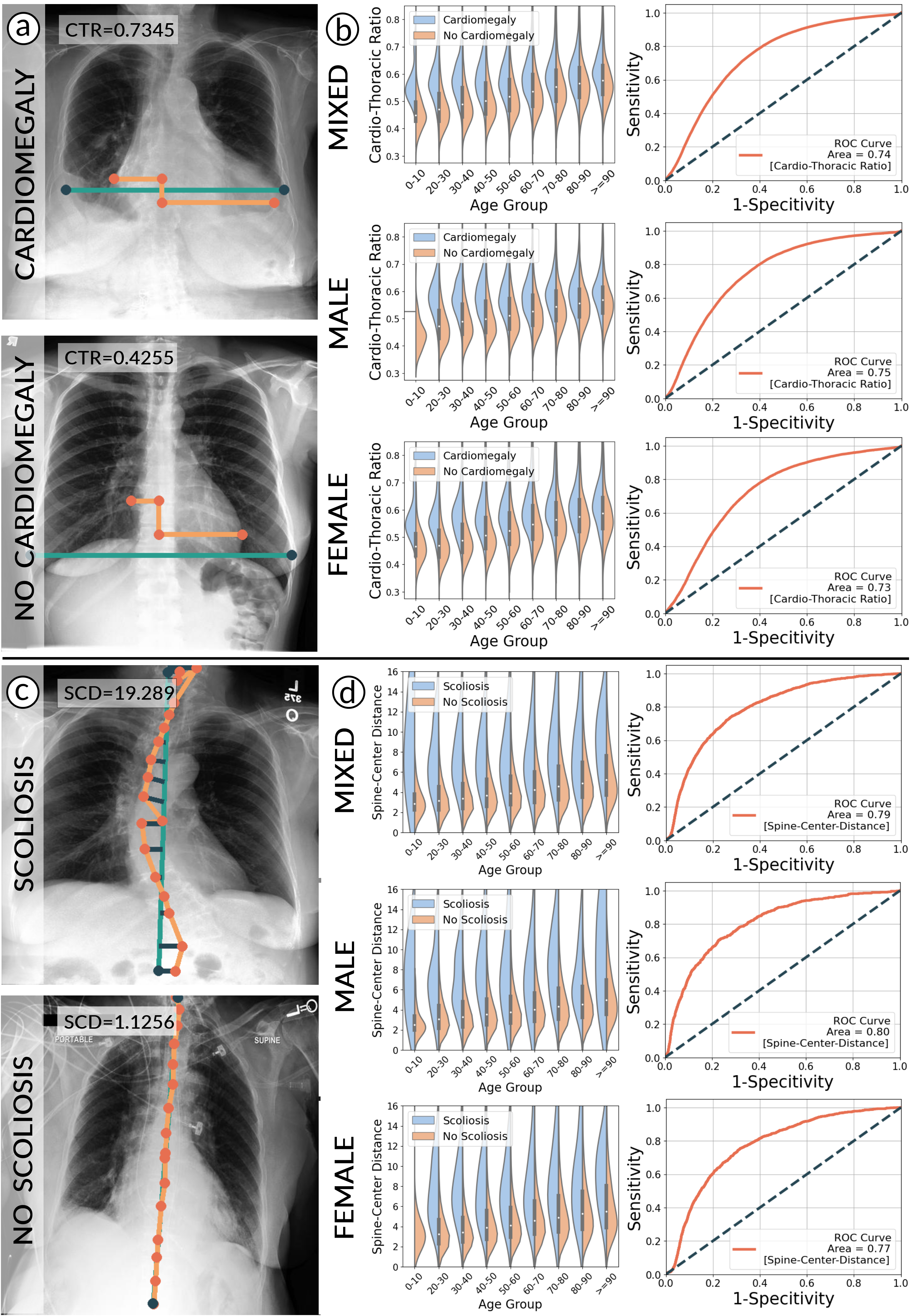}
    \caption{a) qualitative examples of the calculation of the CTR for presence/absence of cardiomegaly. b) We show the CTR distributions of the PA CXR of the MIMIC dataset for sex, pathology and age-group. We further show the predictive value of the CTR for cardiomegaly. c) We show qualitative examples of the calculation of the SCD for presence/absence of scoliosis. d) We show the SCD distributions of the PA CXR of the MIMIC dataset for sex, pathology and age-group. We further show the predictive value of the SCD for scoliosis. }
    \label{fig:extractor}
\end{figure}
We use all PA CXR from the MIMIC dataset and use the publically available automatically pre-extracted labels from medical reports. We extract 41,633 images presenting cardiomegaly and 189,374 images stating either no cardiomegaly or no indication in such a way. 
We infer our CXAS model from all images and extract the cardio-thoracic ratio (CTR) via the lung and heart segmentations. For the calculation of the CTR, we follow Caffey et al.~\cite{caffey1978pediatric}

To infer the difference in this value between pathological and non-pathological cases, we conduct a t-test. Furthermore, to identify the diagnostic ability of this value, we display ROC curves for different subsets of the dataset according to sex.

In Fig. \ref{fig:extractor}$\cir{a}$, we show qualitative examples of the calculation of the cardio-thoracic-ratio for patients presenting cardiomegaly and not presenting cardiomegaly. For the case of the pathology, we see that the border of the heart is expanded,  leading to a CTR above 0.7, whereas, for the absence case, the CTR is 0.4255. 

When conducting a t-test, we get a value of 139.71
(p-value $>$ 0.0001) indicating a strong difference in CTR between positive and negative patients for cardiomegaly.
We show this property in the violin plots on the left of Fig. \ref{fig:extractor}$\cir{b}$, highlighting the CTR distributions for sex, age group, and pathology. Typically, the CTR  for the presence and absence of cardiomegaly has noticeable shifts, with both respective means increasing with age. This average increase in CTR, however, becomes less in the age group of $70 - \geq 90$, with the disparity in mean CTR between cardiomegaly and no cardiomegaly reducing the older a patient gets. This result, in turn, lessens the insight of CTR with age. We observe this property for male and female patients. 
On the right of Fig. \ref{fig:extractor}$\cir{b}$, we show the ROC-Curve for pathology classification based on the CTR score. We see a difference in the diagnostic performance of the CTR for female and male patients with an AuROC of 0.73 and 0.75.

\subsubsection*{Automated Identification of the Center-Spine Distance via CXAS}
We use all PA CXR from the MIMIC dataset and straightforwardly filter medical reports for the term \textit{scoliosis}. We extract 2,054 images presenting scoliosis and 228,953 images not containing this phrase. 
We infer our CXAS model from all images and propose a spine-center distance (SCD) to measure the shift of the individual vertebrae. We calculate this by using the center point of the individual vertebrae segmentations. We then regress a centerline based on these vertebrae centers and compute the sum over the minimal distances of all vertebrae centers to the centerline. 

To infer the difference in the SCD value between pathological and non-pathological cases, we conduct a t-test. Furthermore, to identify the diagnostic ability of this value, we display ROC curves for different subsets of the dataset according to sex.

In Fig. \ref{fig:extractor}$\cir{a}$, we show qualitative examples of the calculation of the SCD for patients for the presence and absence of scoliosis. We can see that the vertebrae segmentation presents a noticeable curvature for the scoliosis case, leading to an SCD above 19, while for the patient with a healthy spine, the SCD is 1.125. 

When conducting a t-test, we get a value of 35.30 
(p-value $<$ 0.0001), indicating a statistically significant difference in SCD between positive and negative patients for scoliosis.
On the left of Fig. \ref{fig:extractor}$\cir{b}$, we show the distributions of the SCD in the form of violin plots for sex, age group, and pathology. 
Similarly to the CTR, SCD increases with age for the presence and absence of scoliosis. The results show that the older the patient gets, the less the disparity in mean SCD between scoliosis and no scoliosis becomes, indicating that the SCD becomes slightly less insightful the older the patient is. However, different from the CTR, the standard deviation of the SCD is considerably higher, and there is a relatively larger difference in means, leading to a higher predictive value for the identification of scoliosis.

On the right of Fig. \ref{fig:extractor}$\cir{b}$, we show the ROC-Curve for pathology classification based on the SCD score. We see a difference in the diagnostic performance of the SCD for female and male patients with an AuROC of 0.77 and 0.80. 
\section*{Limitations}

Our method is not without its limitations. While accurate target masks can be generated, the dataset size is limited by the amount and characteristics of the available CT scans.  CTs are typically a much scarcer resource than the common CXR, making it difficult to scale this approach to the levels of natural image datasets with millions of images due to the lack of volumetric data in comparison. CXR also tends to use a harder type of radiation than standard CTs. This fact, paired with contrast agents in CTs, might lead to a divergence of visual quality between the projected X-rays and real ones. Furthermore, the generated masks will not reach the resolution without running into minor interpolation errors as our projections enable a segmentation at the size of $512\times512$, common CXR approach resolutions of $2048\times2048$.

Due to the dataset properties of PAX-Ray++, which we expand on in Methods, the resulting models are not as robust to novel views or irregular orientations such as heavy rotations.

We, however, argue that through the use of stronger models, enhanced heavy data augmentations and further post-processing rules these errors can be minimized.

\section*{Discussion}
Anatomical segmentation in CXR is a challenging task as manual annotation of large datasets is nearly impossible due to superimposed  structures in the imaging domain. We take advantage of the fact that the CT and X-Ray imaging domains are reconstructed from similar sources of ionizing radiation, allowing us to collect anatomical annotations from various sources in the pixel-wise distinct CT domain, which we subsequently project to a 2D plane resembling an annotated X-Ray. 

Our proposed workflow aggregates label sources from ten different source datasets and utilizes strong 3D segmentation models to generate anatomically grounded pseudo-labels for 7377 thorax volumes. Using those, we create the PAX-Ray++ dataset, the first dataset with dense anatomical annotations in the CXR domain spanning label categories from abdominal and mediastinal regions to specific bones and organs. This dataset contains more than 14 thousand images and is, with more than 2 million instances, comparable in size to natural image datasets such as MS COCO. 

This procedure allows us to develop the first fine-grained anatomy segmentation models for chest X-Rays. By training on the PAX-Ray++ dataset, we create CXAS models that perform well on projected CTs and show a high agreement with medical personnel on CXR data. 

We validate our method on projected data, which shows good performance cross all classes. It can be noted that classes in the lateral view tend to have slightly better scores than their frontal counterparts, which is also attributed to the fact that frontal classes tend to be more fine-granular (i.e., left/right split).
Our experiments also show on regular chest X-rays good alignment with trained experts. It is noted that even trained experts' annotations do not always agree in this domain, especially for abdominal classes. 

We demonstrate how to generate meaningful and interpretable patient features by extracting custom information from anatomical segmentations. We show how these can help in the large-scale analysis of patient cohorts for the exemplary pathologies 'cardiomegaly' and 'scoliosis'. We can identify age-based trends for these cases and highlight how these anatomy-based features aid the interpretation of a patient. 

We believe that anatomy segmentation can aid the development of medical reports by providing grounded information. These segmentations can similarly enhance pathology recognition models by identifying associated regions and enabling potential reasoning approaches stemming from the co-occurrence of abnormal visual patterns and anatomy.

\bibliography{sn-bibliography}
\section*{Methods}
\subsection*{Comparison of X-Ray and CT}

X-Ray imaging is a basis for several non-invasive imaging techniques that use ionizing radiation in the form of X-Rays to produce images of the body's internal structures. When X-Rays pass through the body, dense tissues, such as bone, absorb more radiation than softer tissues, such as muscle or fat. This difference in absorption allows X-Rays to depict the body's internal structures.

In conventional chest X-Ray exams, the erect patient is positioned between an X-Ray tube and a detector. The collected data by the detector is then used to create a comprehensive image. The main distinctions in positioning (such as anteroposterior (AP) and posteroanterior (PA), both of which we refer to a \textit{Frontal}, or lateral (L)), are related to which view enables the most insight. 

In a similar manner, CTs are an imaging technique building on ionizing radiation to produce detailed images of the body's internal structures. However, instead of using a fixated X-Ray tube and detector as with X-Rays, CT scanners use multiple X-Ray sources and detectors that rotate around the body, producing a series of cross-sectional images to compute a slice.
This process is done until all slices of a desired body part are taken, resulting in a detailed three-dimensional volume. This eliminates superimposition common in standard X-Rays, enabling a radiologist to more easily interpret a patient.
While the CT scan enables better diagnostic possibilities than the X-Ray, the X-Ray is more accessible and it involves lower radiation doses. 

 Often, an X-Ray is one of the first steps in the radiologic diagnostic process.
If not enough information to diagnose a particular condition can be extracted, a CT scan may be ordered. Hence, it is crucial to enable systems to extract the maximum information to avoid unnecessary procedures.

\subsection*{Projecting CT back to X-Ray }\label{sec:panda_projection}
\renewcommand{\arraystretch}{0.65}
\setlength{\tabcolsep}{2pt}
\setlength{\abovecaptionskip}{6pt}
\setlength{\belowcaptionskip}{-10pt}

\begin{figure}[t!]
    \centering
    \begin{tabular}{c ccc || ccc}
    \toprule
    & \multicolumn{3}{c}{OpenI (X-Ray)} & \multicolumn{3}{c}{PAXRay++ (Projected)} \\
    \midrule
    \multirow{3}{*}{\rotatebox{90}{Frontal\phantom{000}}} & \includegraphics[width=0.15\linewidth,height=0.15\linewidth]{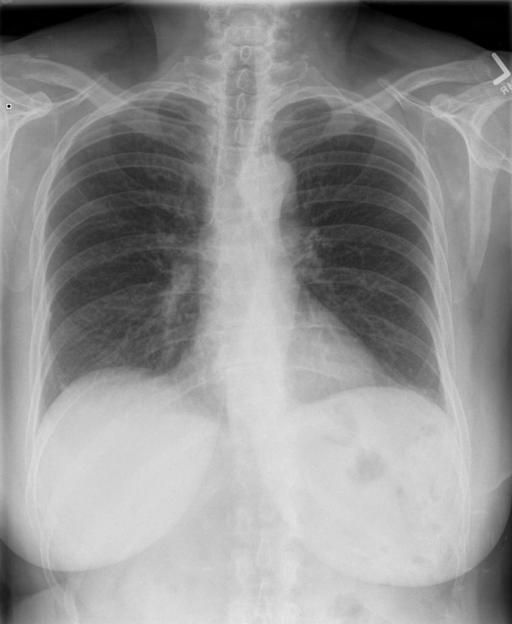}& \includegraphics[width=0.15\linewidth,height=0.15\linewidth]{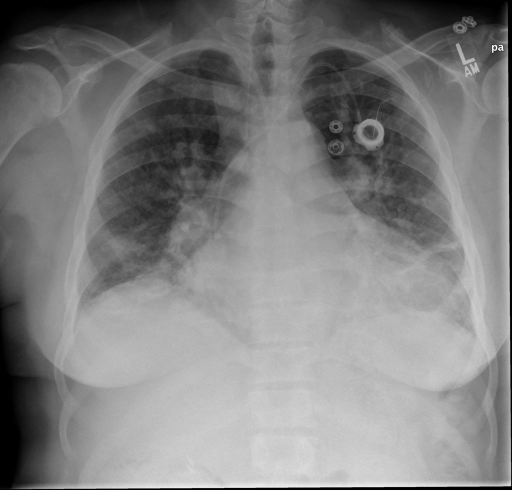}& \includegraphics[width=0.15\linewidth,height=0.15\linewidth]{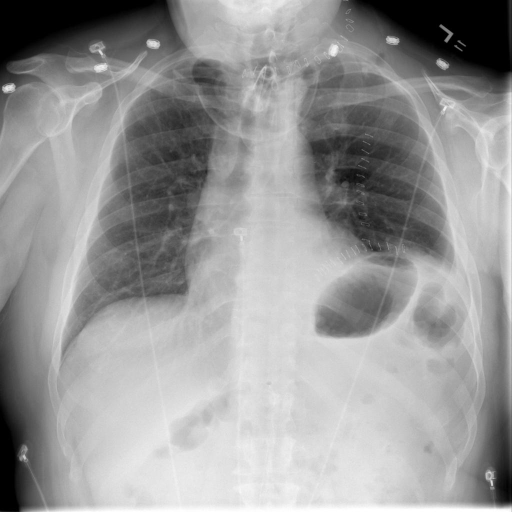}& 
    \includegraphics[width=0.15\linewidth,height=0.15\linewidth]{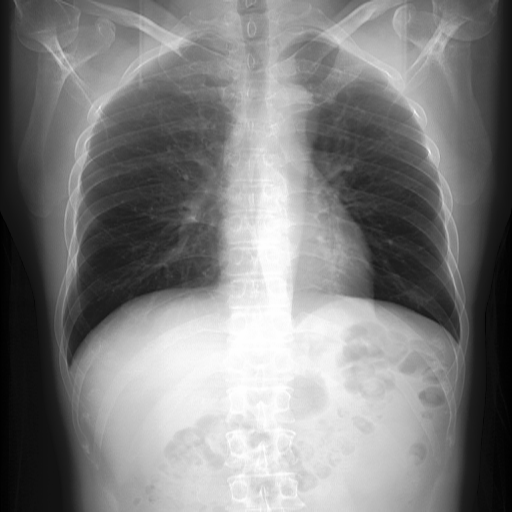}& \includegraphics[width=0.15\linewidth,height=0.15\linewidth]{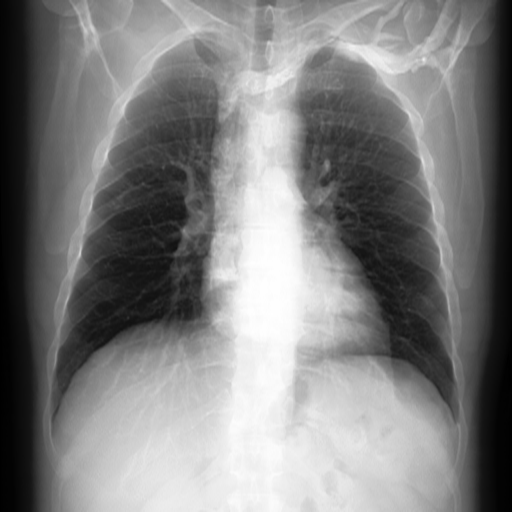}& \includegraphics[width=0.15\linewidth,height=0.15\linewidth]{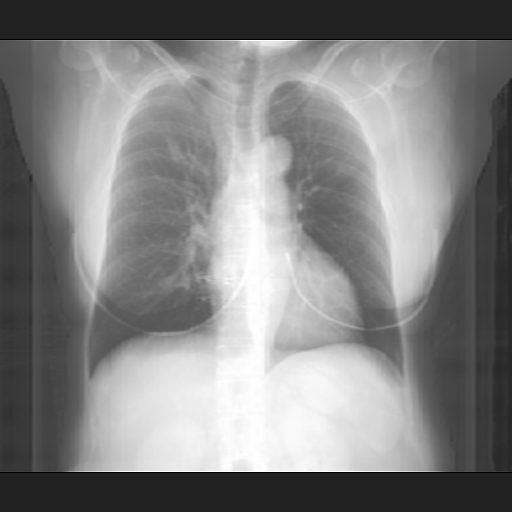} \\
    & \includegraphics[width=0.15\linewidth,height=0.15\linewidth]{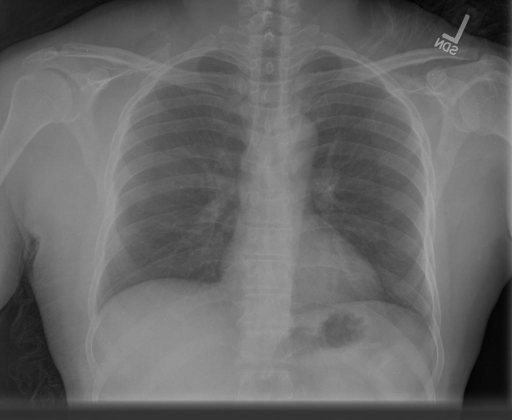}& \includegraphics[width=0.15\linewidth,height=0.15\linewidth]{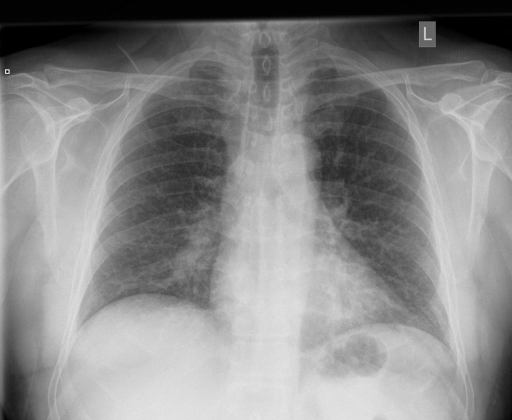}& \includegraphics[width=0.15\linewidth,height=0.15\linewidth]{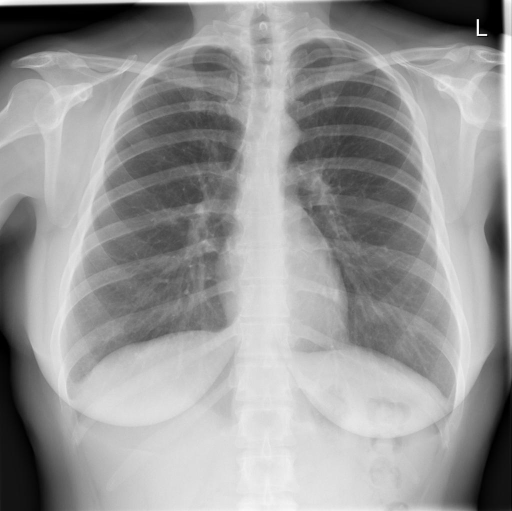}& 
    \includegraphics[width=0.15\linewidth,height=0.15\linewidth]{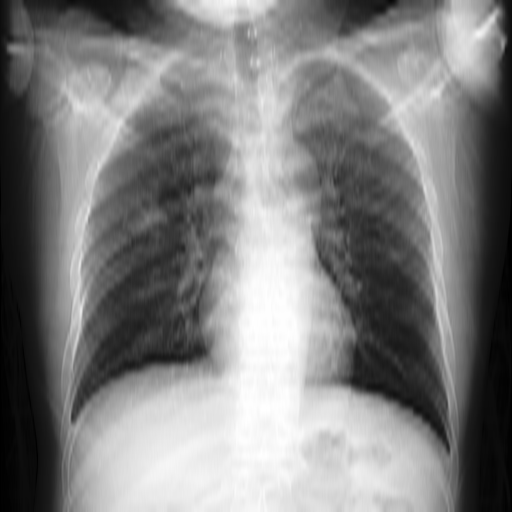}& \includegraphics[width=0.15\linewidth,height=0.15\linewidth]{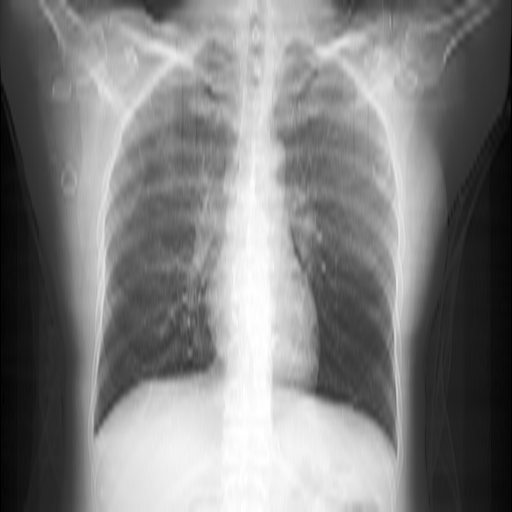}& \includegraphics[width=0.15\linewidth,height=0.15\linewidth]{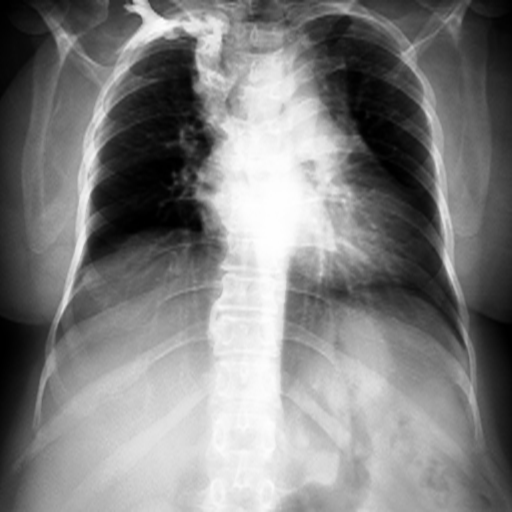}\\\midrule\midrule
    \multirow{3}{*}{\rotatebox{90}{Lateral\phantom{000}}}& \includegraphics[width=0.15\linewidth,height=0.15\linewidth]{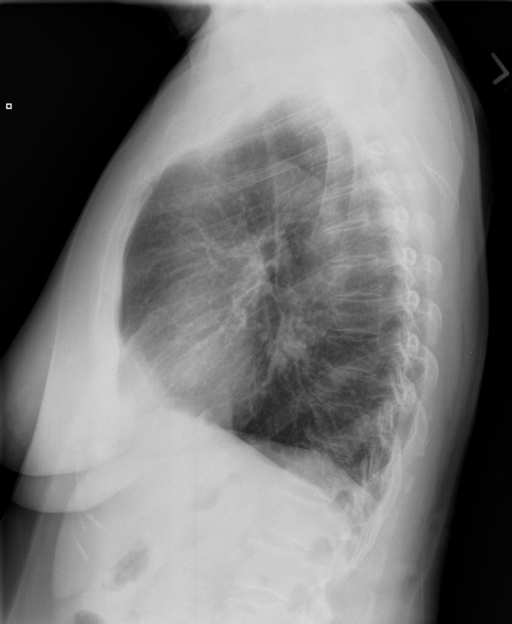}& \includegraphics[width=0.15\linewidth,height=0.15\linewidth]{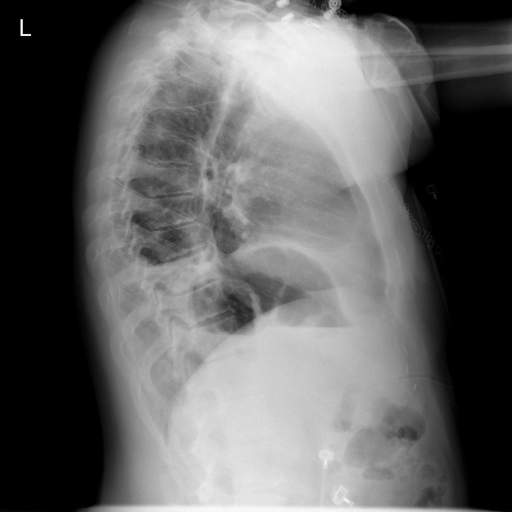}& \includegraphics[width=0.15\linewidth,height=0.15\linewidth]{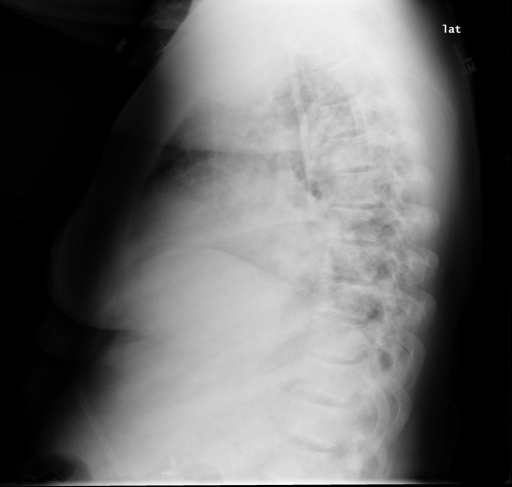}& 
    \includegraphics[width=0.15\linewidth,height=0.15\linewidth]{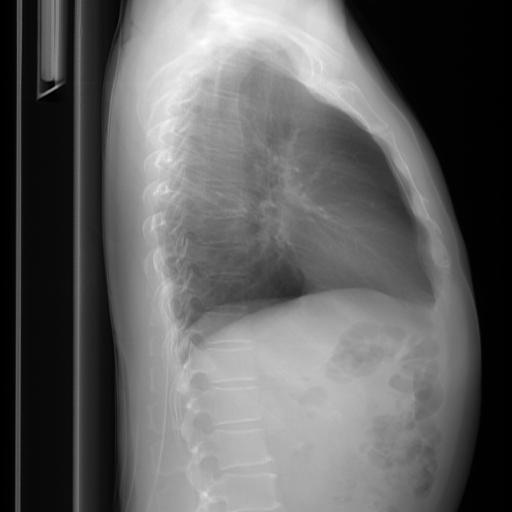}& \includegraphics[width=0.15\linewidth,height=0.15\linewidth]{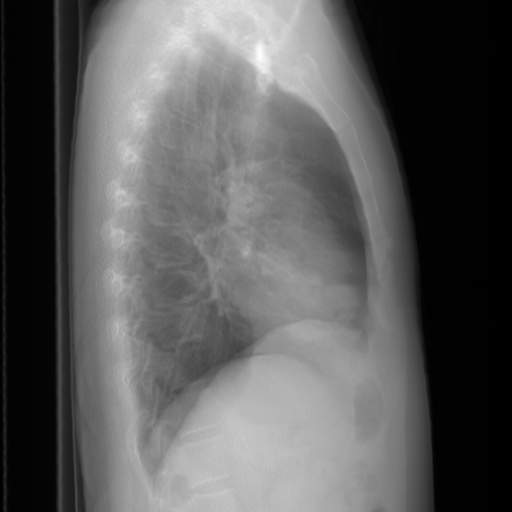}& \includegraphics[width=0.15\linewidth,height=0.15\linewidth]{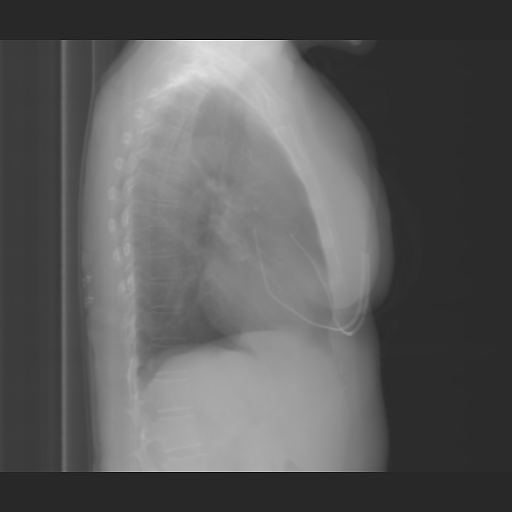} \\
    & \includegraphics[width=0.15\linewidth,height=0.15\linewidth]{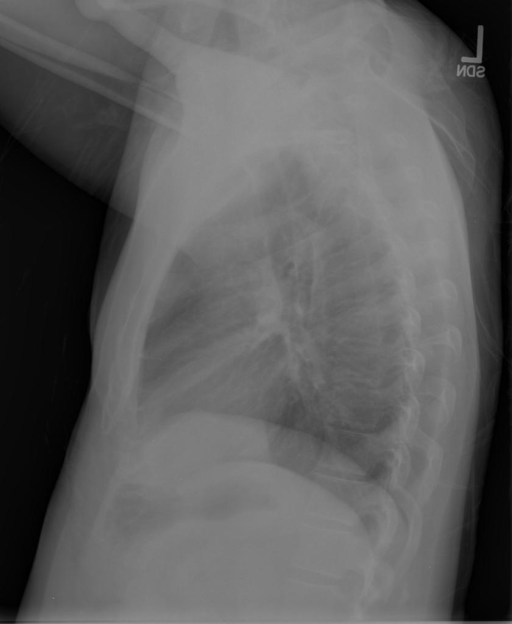}& \includegraphics[width=0.15\linewidth,height=0.15\linewidth]{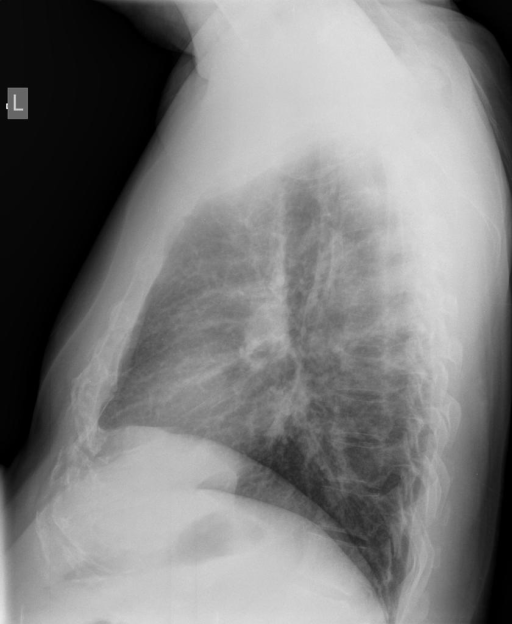}& \includegraphics[width=0.15\linewidth,height=0.15\linewidth]{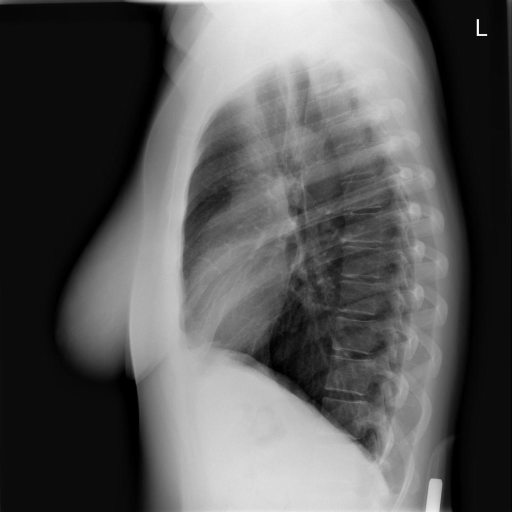}& 
    \includegraphics[width=0.15\linewidth,height=0.15\linewidth]{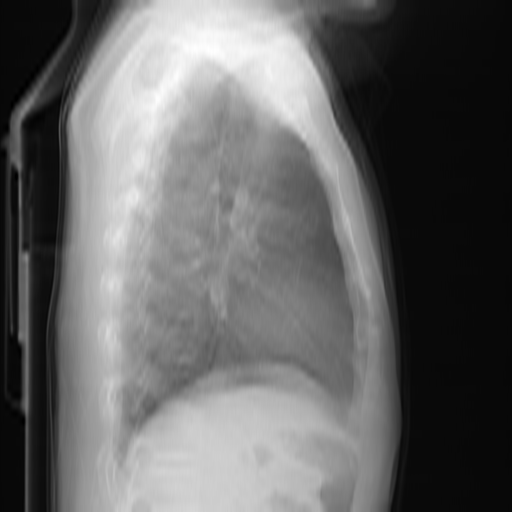}& \includegraphics[width=0.15\linewidth,height=0.15\linewidth]{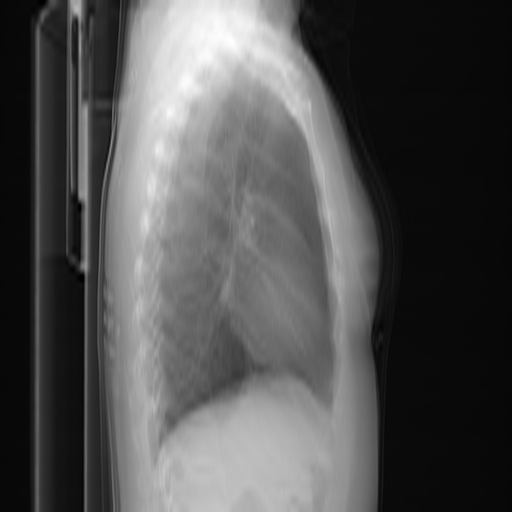}& \includegraphics[width=0.15\linewidth,height=0.15\linewidth]{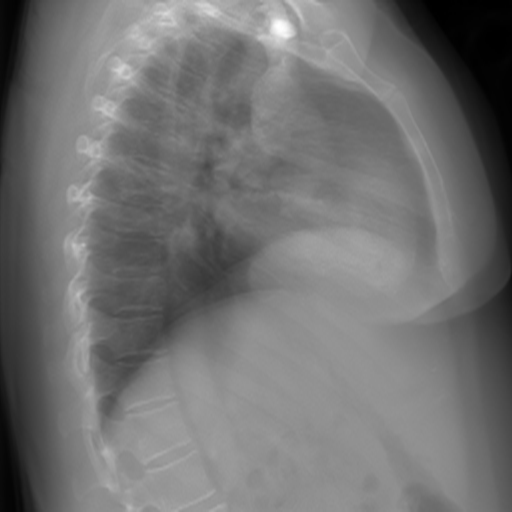}\\
    \bottomrule
    \end{tabular}
    \caption{Comparison of different real and projected samples for the frontal and lateral view}
    \label{fig:syn_real}
\end{figure}

The back-projection is related to the concepts of Mean Intensity Projection and Maximum Intensity Projection~\cite{fishman2006volume}, which allow one to gain quick insights on a patient without requiring a thorough look through the CT. Unlike in the Mean- or Maximum-Intensity-projection, we want to generate images that look visually similar to real X-Rays.

We start the CT projection process by extending Matsubara~\etal's~\cite{matsubara2019generation} aproach. Instead of using the volume directly, we first collect the outline of the patient to separate it from the table he is lying on. For this, we clip the CT volume to the standard 12-bit range. We get the body mask via first thresholding at a value of $-100$, filling each contour and extracting the connected component of the highest volume. We found that utilizing the additional body masks assists with excluding unwanted structures.  
Afterwards, we collect the bone structures  through slicewise generalized histogram thresholding~\cite{barron2020generalization}. We do not use our segmentation masks for bone extraction as some labels such as the humerus are not included and we instead want to provide a complete projection of the human body. 
The bone and body volumes are added up with a weight of $1$ and $0.3$
 We then average the volume along the coronal plane to retrieve a frontal projection along the sagittal plane for a lateral one. We rescale all image projections to 8-bit via min-max feature scaling and resize them to $512\times512$ using Lanczos interpolation.

For the projection, we utilize Python 3.10 and packages such as NumPy~\cite{harris2020array}, SciPy~\cite{2020SciPy-NMeth} and SimpleITK. 

We compare our projected X-Rays with samples from the OpenI dataset for frontal and lateral views in Figure~\ref{fig:syn_real}. The differences in the frontal view are due to the different positioning of the shoulder girdle. In the X-Rays, the arms are usually placed alongside the body, while in the projected images, the arms are raised due to the nature of the CT scan. 
In the lateral view, the X-Rays show a more comprehensive range of orientation and pose. However, the projected images, typically taken while the patient is lying down, result in similar poses between the different images. 
This leads to visual differences between images of female patients in both frontal and lateral views, 
such as the third column and second row of real X-Ray images and the first column and 
second row of projected images.

\subsection*{Label Sources of PAX-Ray++}

\subsubsection*{Generation of Volumetric Labels}
\begin{figure*}[t]
\centering
    \begin{tabular}{c}
        \includegraphics[width=\linewidth]{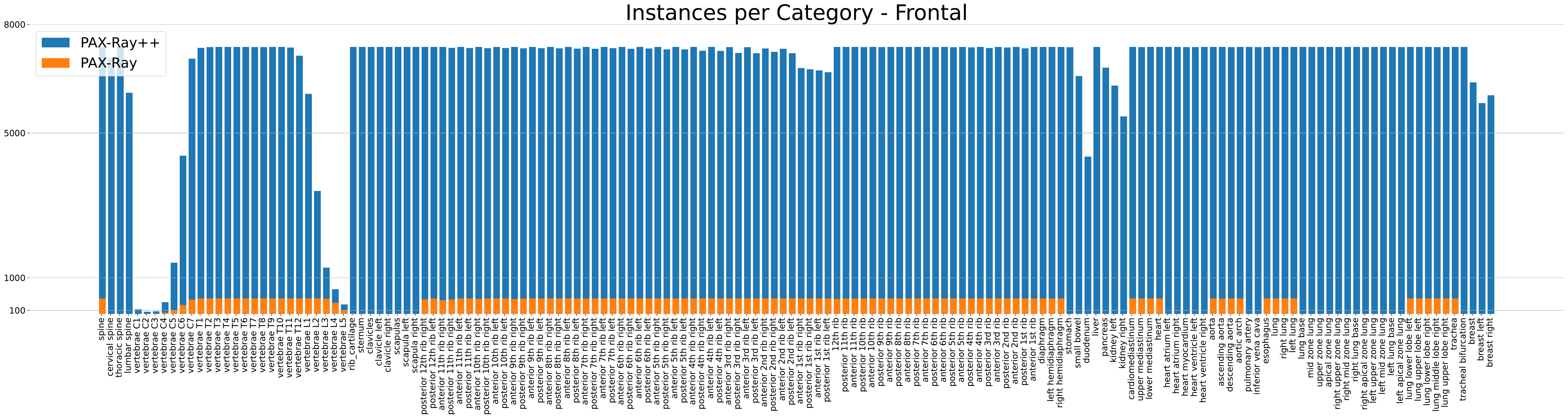}  \\ 
        \includegraphics[width=\linewidth]{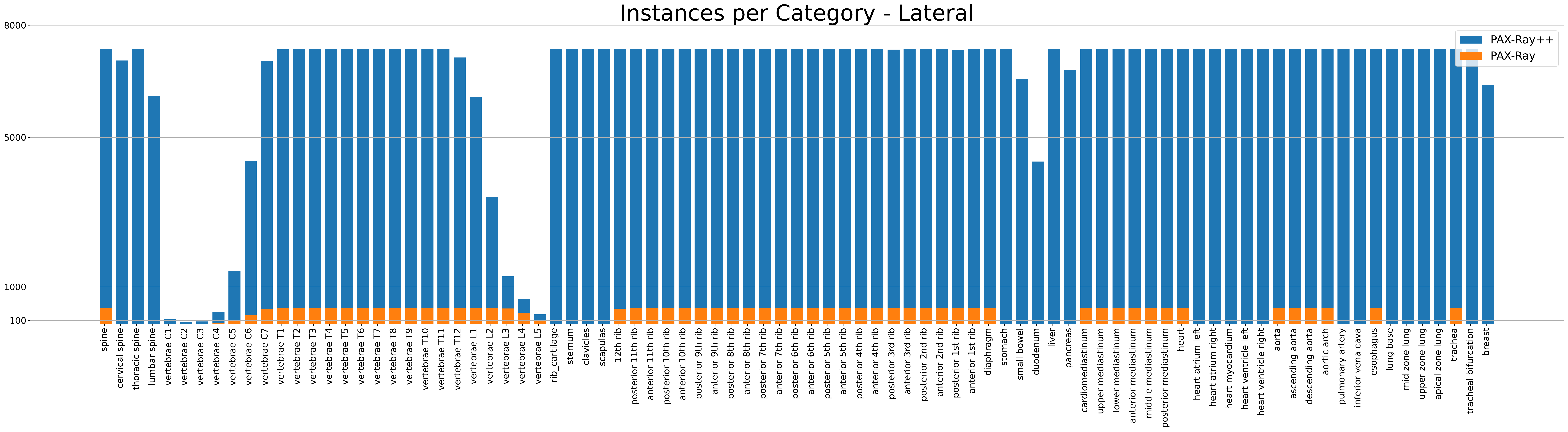}
    \end{tabular}
    \caption{Label distribution comparison between the PAX-Ray and PAX-Ray++ dataset.}
    \label{fig:labelstructure_extendedpaxray}
\end{figure*}

We consider datasets focusing on different body systems and available segmentation tools to provide a more leveled label distribution. We use these to train five folds of full resolution nnUNets~\cite{isensee2021nnu} and keep a separate set for validation purposes. For inference, we use an ensemble of all folds with test time augmentation, such as horizontal flipping.
We list the source datasets, their volume information, and the segmentation performance of their resulting networks in Table~\ref{tab:ch5_datasets_overview}. 

\begin{enumerate}
    \item The Pediatric dataset~\cite{jordan2022pediatric} contains 359  volumes of pediatric patients, with 29 labels spanning the entire body. We discard classes such as 'skin' which bear no relevance for the final task. The segmentation model achieved an IoU of 0.833. 
    \item The used Total Segmentator dataset~\cite{wasserthal2022totalsegmentator} includes 727 volumes, with 104 labels for whole body segmentation, including classes such as humerus and the atria of the heart. We achieve an IoU of 0.9251.
    \item The SegTHOR dataset~\cite{lambert2020segthor} includes 40 volumes, with 5 labels for thoracic organs at risk. We achieve an IoU of 0.9232.
    \item The PARSE dataset~\cite{parse} includes 100/30 volumes for pulmonary artery segmentation. On the challenge benchmark, we achieve an IoU of 0.751. 
    \item  The BTCV dataset~\cite{malabdomen} includes 30/20 images, with 14 labels for abdominal segmentation, on which we achieve an IoU of 0.8511.
    \item The dataset by Ma et al. ~\cite{ma2021abdomenct} includes 50 images, with 13 labels for abdominal segmentation. With an IoU of 0.9093. 
    \item The RibSeg dataset~\cite{yang2021ribseg} includes 580 volumes. While the original annotations segment the ribs as a single class along the outskirts of the ribs, we fill the ribs via thresholding\cite{barron2020generalization} and morphological operations and subsequently partition these via connected components into the 12 left and right ribs, with 24 labels for rib segmentation with an IoU of 0.8996.
    \item The Verse dataset~\cite{sekuboyina2021verse} includes 300 volumes, with 23 labels for vertebrae segmentation. We achieve an IoU of 0.8337. 
\end{enumerate}

We further utilize pre-trained classification models such as the Body Composition Analysis by Koitka~\etal~\cite{koitka2021fully} and the lung lobe segmentation by Hoffmaninger~\etal~\cite{ltrc}.
\setlength{\tabcolsep}{3.5pt}

\begin{table*}[t]
    \centering
    \tiny
    \begin{tabular}{cccccccc}
    \toprule
    Source & Series & Labels & Label Domain & Spacing & Slices & IoU \\
    \midrule
             Pediatric~\cite{jordan2022pediatric} & $ 287 / 72 $ & 29 & WholeBody & [$  0.52 \pm 0.11 , 0.52 \pm 0.11 , 1.76 \pm 0.52  $]  & $ 306.6 \pm 245.4 $ & $0.833$ \\
         Total Segmentator~\cite{wasserthal2022totalsegmentator} & $ 581 / 146 $ & 104 & WholeBody & [$  1.5 \pm 0.0 , 1.5 \pm 0.0 , 1.5 \pm 0.0  $]  & $ 259.0 \pm 130.3 $ & $0.9251$\\
         
         SegTHOR~\cite{lambert2020segthor} &  $ 40 / 20 $ & 5 & Thoracic & [$  1.00 \pm 0.09 , 1.00 \pm 0.09 , 2.39 \pm 0.23  $]  & $ 184.7 \pm 30.35 $ & $0.9232$ \\
         PARSE~\cite{parse} & $ 100 / 30 $ & 1 & P. Artery & [$  0.67 \pm 0.07 , 0.67 \pm 0.07 , 0.99 \pm 0.01  $]  & $ 301.4 \pm 31.22 $ & $0.751$\\
         BTCV~\cite{malabdomen} & $ 30 / 20 $ & 14 & Abdominal & [$  0.78 \pm 0.09 , 0.78 \pm 0.09 , 3.87 \pm 1.00  $]  & $ 123.3 \pm 32.19 $ & $0.8810$ \\
         Ma et al~\cite{ma2021abdomenct} & $ 40 / 10 $ & 13 & Abdominal & [$  0.81 \pm 0.07 , 0.81 \pm 0.07 , 2.63 \pm 0.52  $]  & $ 95.88 \pm 9.000 $& $0.9093$\\
        RibSeg~\cite{yang2021ribseg} & $ 420 / 160 $ & 24 & Ribs & [$  0.74 \pm 0.07 , 0.74 \pm 0.07 , 1.13 \pm 0.14  $]  & $ 359.5 \pm 59.06 $ & $0.8996$\\
        Verse~\cite{sekuboyina2021verse} & $ 112 / 100 $ & 23 & Spine & [$  0.79 \pm 0.23 , 0.79 \pm 0.23 , 1.29 \pm 0.65  $]  & $ 444.8 \pm 348.6 $ & $0.8387$\\

        \midrule
         RSNA-PE \cite{colak2021rsna} & $ 7302 $ & - & -& [$  0.67 \pm 0.08 , 0.67 \pm 0.08 , 1.25 \pm 0.22  $]  & $ 237.6 \pm 54.00 $ & - \\
         RibFrac \cite{ribfrac2020}& $ 660 $ & - &- & [$  0.74 \pm 0.07 , 0.74 \pm 0.07 , 1.13 \pm 0.14  $]  & $ 359.5 \pm 59.06 $ & - \\
          LIDC-IDRI ~\cite{armato2011lung} & $ 1036 $ & - & -& [$  0.68 \pm 0.08 , 0.68 \pm 0.08 , 1.74 \pm 0.83  $]  & $ 237.9 \pm 132.7 $ & - \\
         COVID-19-AR~\cite{desai2020chest} & $ 176 $ & - & -& [$  1.57 \pm 0.79 , 1.34 \pm 0.77 , 1.42 \pm 1.05  $]  & $ 438.8 \pm 178.0 $ & - \\
         COVID-19-1~\cite{harmon2020artificial} & $ 215 $ & - & -& [$  0.77 \pm 0.10 , 0.77 \pm 0.10 , 4.76 \pm 0.88  $]  & $ 68.74 \pm 36.24 $& -  \\
         COVID-19-2~\cite{kassin2021generalized} & $ 632 $ & - & -& [$  0.77 \pm 0.10 , 0.77 \pm 0.10 , 4.72 \pm 0.94  $]  & $ 72.21 \pm 48.63 $ & - \\
         \bottomrule
    \end{tabular}
    \caption{Comparison of considered CT datasets for PAX-Ray++ regarding size, number of labels, label domain, volume spacing, number of slices and segmentation performance of a nnUNet~\cite{isensee2021nnu} in 5-fold cross validation. Top half displays datasets used for label aggregation, while the bottom part was used as imaging source for PAX-Ray++.}
    \label{tab:ch5_datasets_overview}
\end{table*}

\subsubsection*{Generation of Additional 2D Labels}
We integrate anatomical regions by modifying several anatomical classes:
\begin{enumerate}
\item The mediastinum stems from the BCA predictions and is split into upper and lower mediastinum along the T4 vertebra.
\item We differentiate the anterior and posterior mediastinum along the heart segmentation boundary.
\item We create the upper and middle lung regions and bases by splitting along the vertical thirds. The apical lung region is defined by the start of the individual clavicles.
\item We define the tracheal bifurcation as starting from 10 slices above the split of the trachea.
\item We maintain the process of the aorta splits and sub-hemidiaphragm from PAX-Ray\cite{Seibold_2022_BMVC}. 
\end{enumerate}

\begin{figure}[p!]
\begin{tabular}{cccccc}
    \toprule
         & Projection & Lungs & Mediastinum & Bones & Abdominal  \\
         \midrule
         \rotatebox{90}{\phantom{00000}RibFrac}& 
         \includegraphics[width=0.175\linewidth,height=0.175\linewidth]{figures/dataset_samples/RibFrac_231_frontal.png}&
         \includegraphics[width=0.175\linewidth,height=0.175\linewidth]{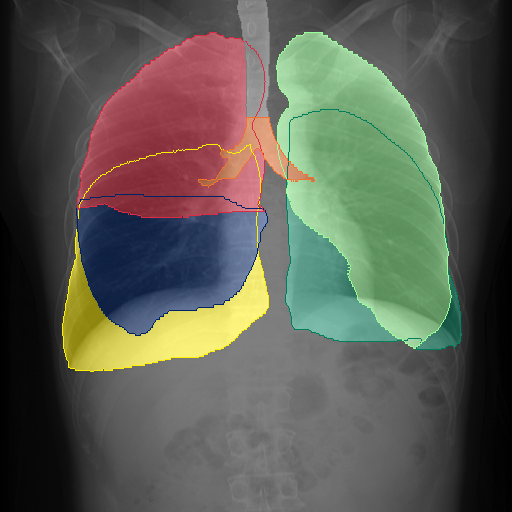}&
         \includegraphics[width=0.175\linewidth,height=0.175\linewidth]{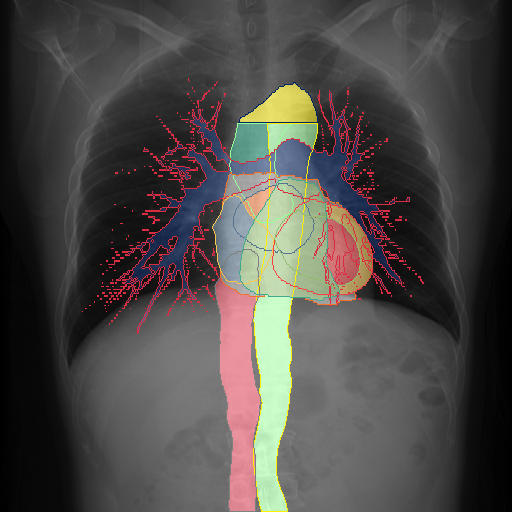}&
         \includegraphics[width=0.175\linewidth,height=0.175\linewidth]{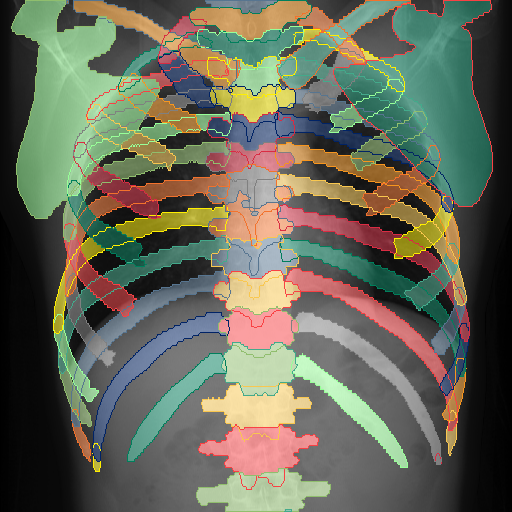}&
         \includegraphics[width=0.175\linewidth,height=0.175\linewidth]{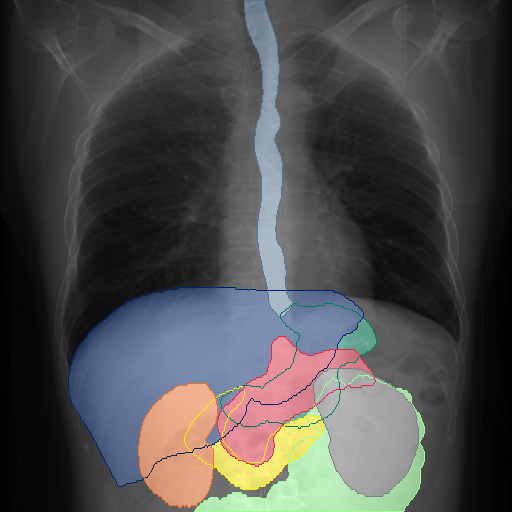}\\
         \rotatebox{90}{\phantom{0000}Covid19AR}& 
         \includegraphics[width=0.175\linewidth,height=0.175\linewidth]{figures/dataset_samples/COVID19AR_4_frontal.png}&
         \includegraphics[width=0.175\linewidth,height=0.175\linewidth]{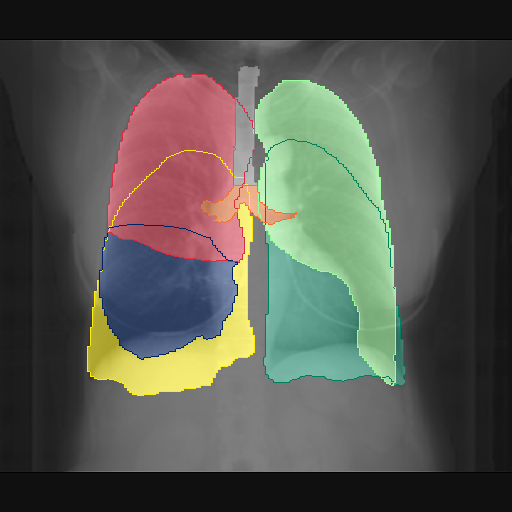}&
         \includegraphics[width=0.175\linewidth,height=0.175\linewidth]{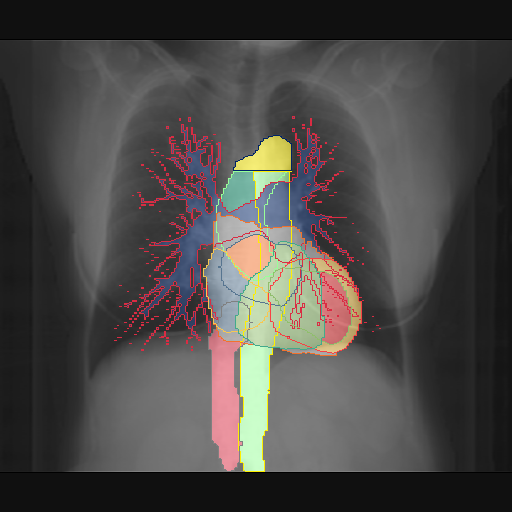}&
         \includegraphics[width=0.175\linewidth,height=0.175\linewidth]{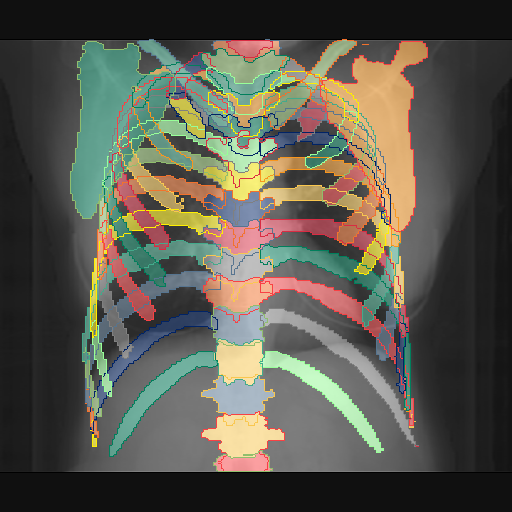}&
         \includegraphics[width=0.175\linewidth,height=0.175\linewidth]{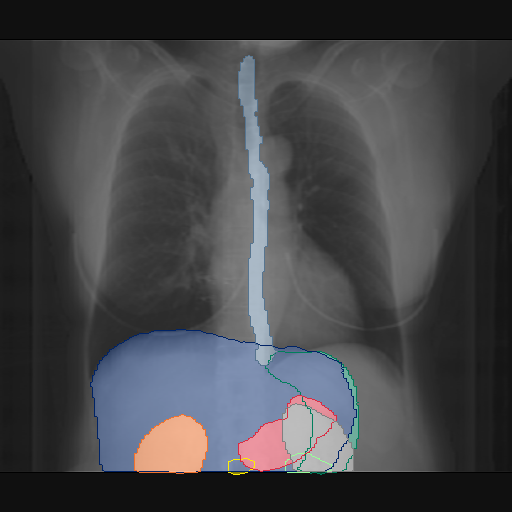}\\
         \rotatebox{90}{\phantom{0000}Covid1910}& 
         \includegraphics[width=0.175\linewidth,height=0.175\linewidth]{figures/dataset_samples/COVID1910_0355_frontal.png}&
         \includegraphics[width=0.175\linewidth,height=0.175\linewidth]{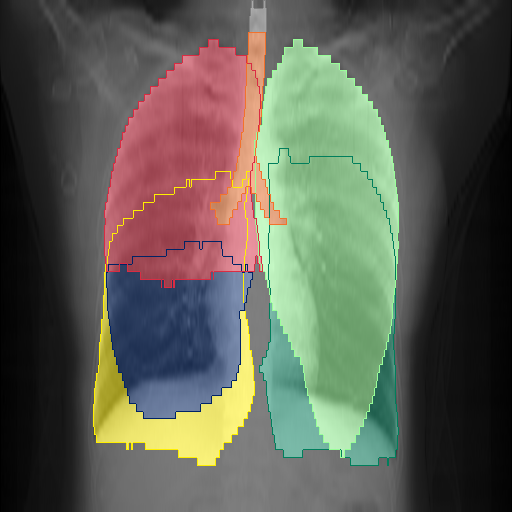}&
         \includegraphics[width=0.175\linewidth,height=0.175\linewidth]{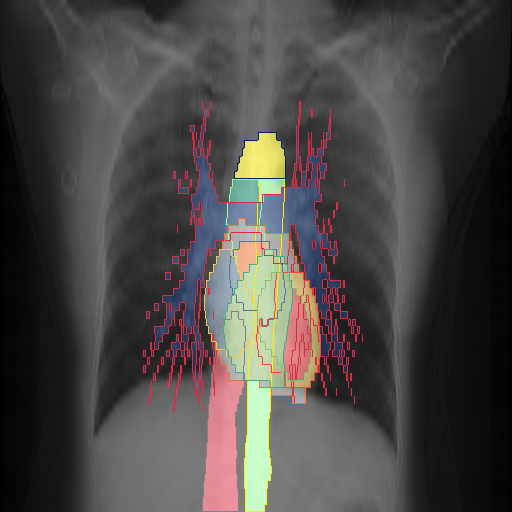}&
         \includegraphics[width=0.175\linewidth,height=0.175\linewidth]{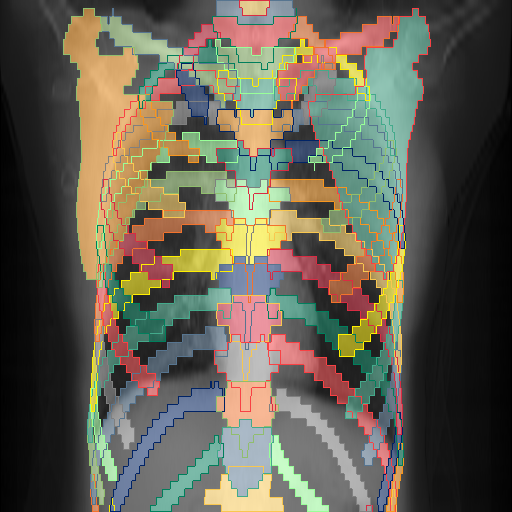}&
         \includegraphics[width=0.175\linewidth,height=0.175\linewidth]{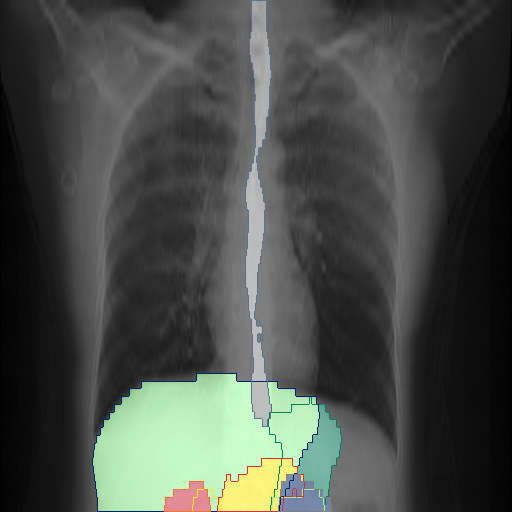}\\
         \rotatebox{90}{\phantom{0000}Covid1920}& 
         \includegraphics[width=0.175\linewidth,height=0.175\linewidth]{figures/dataset_samples/COVID1920_296_frontal.png}&
         \includegraphics[width=0.175\linewidth,height=0.175\linewidth]{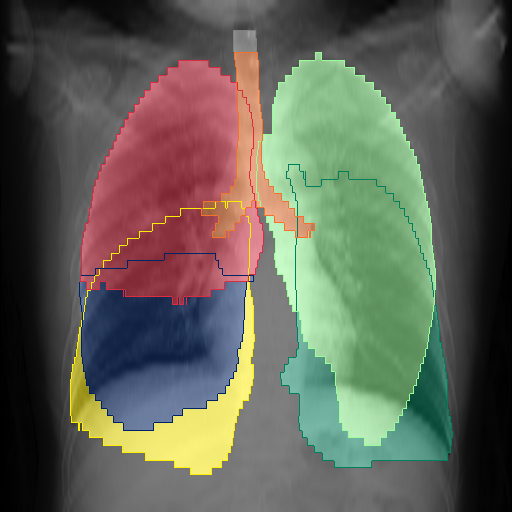}&
         \includegraphics[width=0.175\linewidth,height=0.175\linewidth]{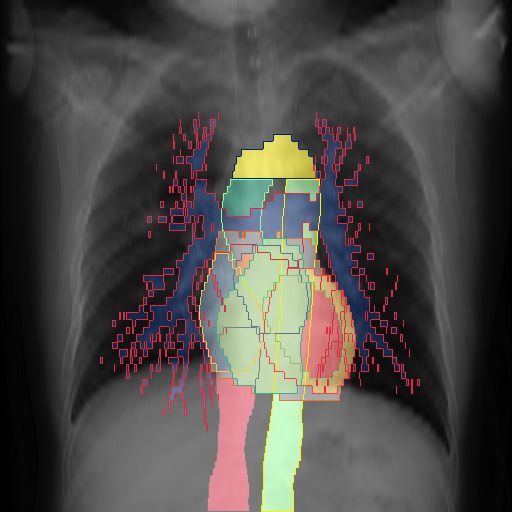}&
         \includegraphics[width=0.175\linewidth,height=0.175\linewidth]{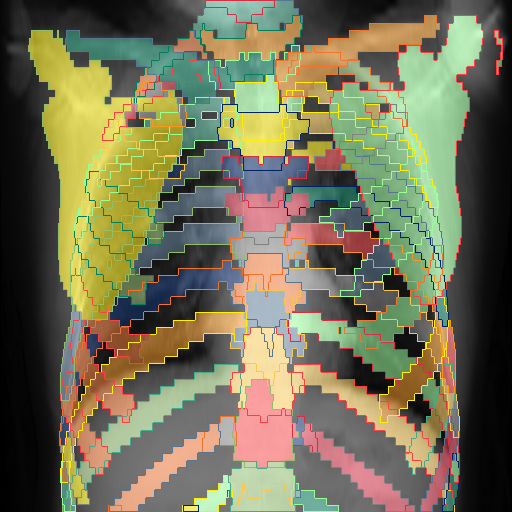}&
         \includegraphics[width=0.175\linewidth,height=0.175\linewidth]{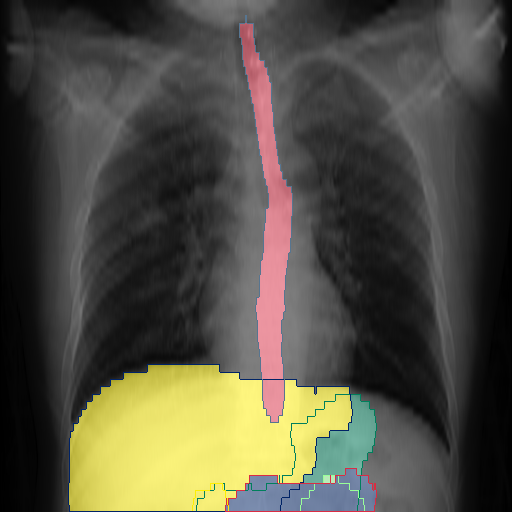}\\
         \rotatebox{90}{\phantom{00000}RSNA PE}& 
         \includegraphics[width=0.175\linewidth,height=0.175\linewidth]{figures/dataset_samples/RSNAPE_fc15f24c6ffb_aa4ad84d4ec6_frontal.png}&
         \includegraphics[width=0.175\linewidth,height=0.175\linewidth]{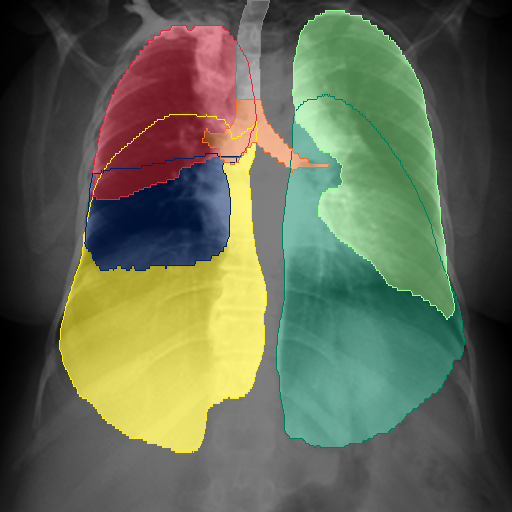}&
         \includegraphics[width=0.175\linewidth,height=0.175\linewidth]{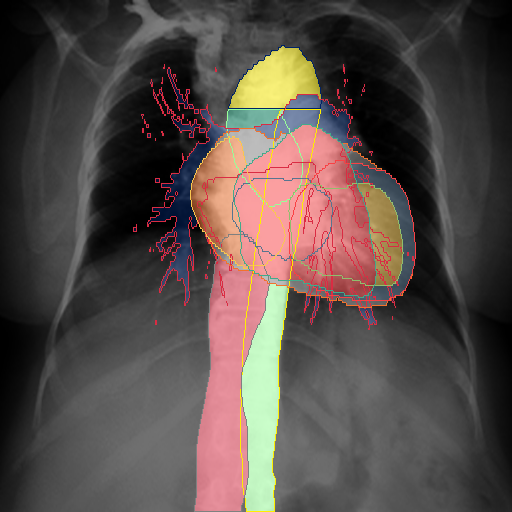}&
         \includegraphics[width=0.175\linewidth,height=0.175\linewidth]{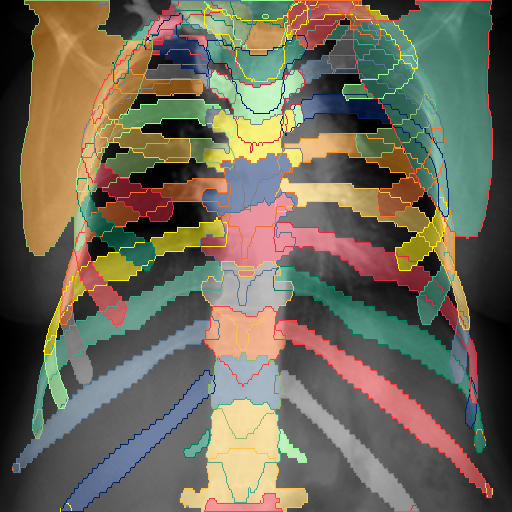}&
         \includegraphics[width=0.175\linewidth,height=0.175\linewidth]{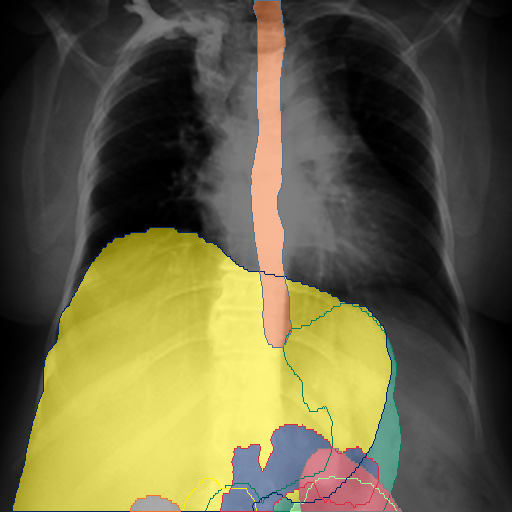}\\
         \rotatebox{90}{\phantom{00}LIDC-IDRI}& 
         \includegraphics[width=0.175\linewidth,height=0.175\linewidth]{figures/dataset_samples/1010b_frontal.png}&
         \includegraphics[width=0.175\linewidth,height=0.175\linewidth]{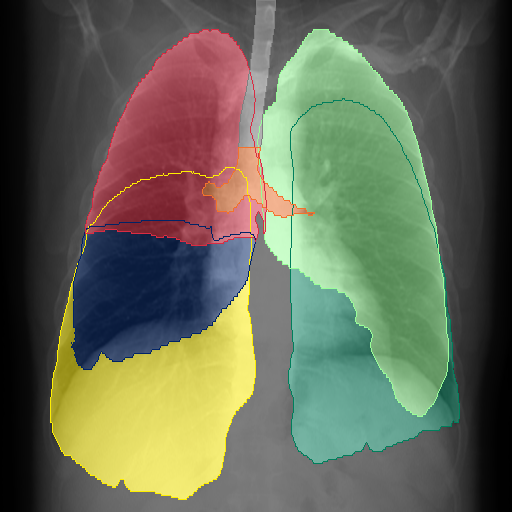}&
         \includegraphics[width=0.175\linewidth,height=0.175\linewidth]{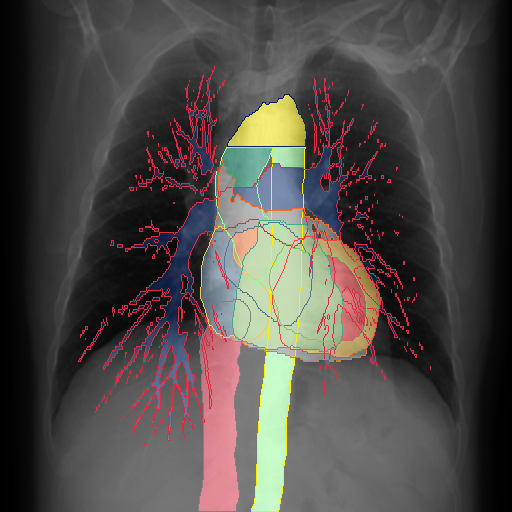}&
         \includegraphics[width=0.175\linewidth,height=0.175\linewidth]{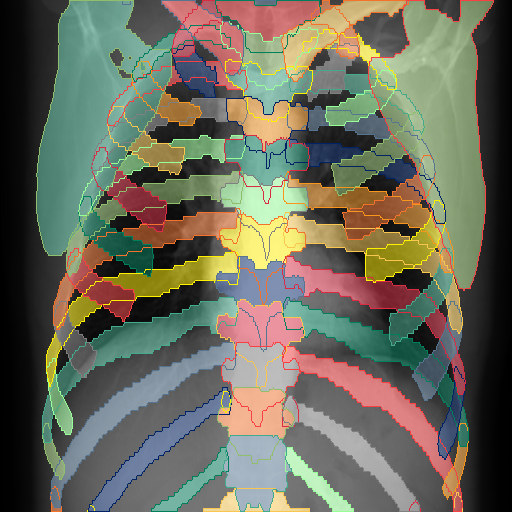}&
         \includegraphics[width=0.175\linewidth,height=0.175\linewidth]{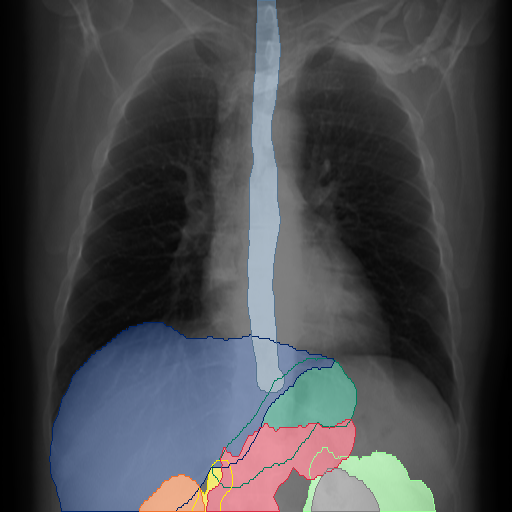}\\
         \bottomrule
    \end{tabular}
    \caption{Sample of frontal projected x-rays from the RibFrac, COVID-19-AR, COVID-19-1, COVID-19-2, RSNA PE, and LIDC-IDRI dataset. We show labels belonging to the categories \textit{Respiratory System}, \textit{Vascular System}, \textit{Bones}, and \textit{Abdomen}.}\label{fig:paxray_frontal}
\end{figure}

\begin{figure*}[p!]
    \begin{tabular}{cccccc}
    \toprule
         \toprule
         & Projection & Lungs & Mediastinum & Bones & Abdominal  \\
         \midrule
         \rotatebox{90}{\phantom{00000}RibFrac}& 
         \includegraphics[width=0.175\linewidth,height=0.175\linewidth]{figures/dataset_samples/RibFrac_231_lateral.png}&
         \includegraphics[width=0.175\linewidth,height=0.175\linewidth]{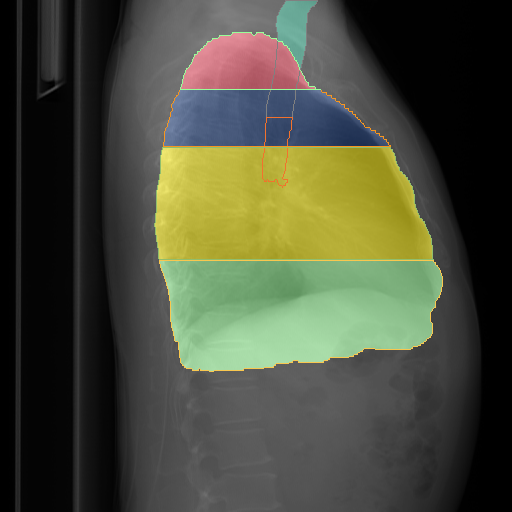}&
         \includegraphics[width=0.175\linewidth,height=0.175\linewidth]{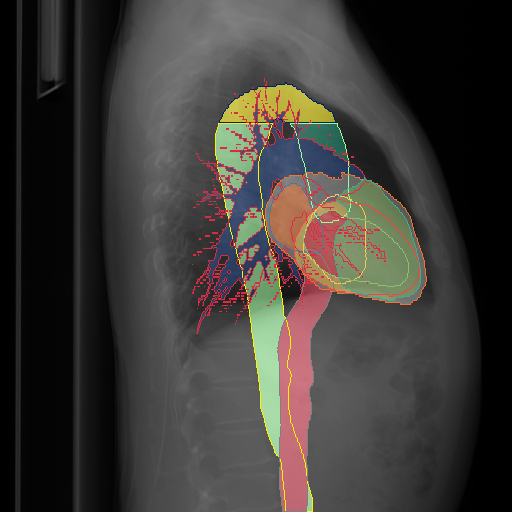}&
         \includegraphics[width=0.175\linewidth,height=0.175\linewidth]{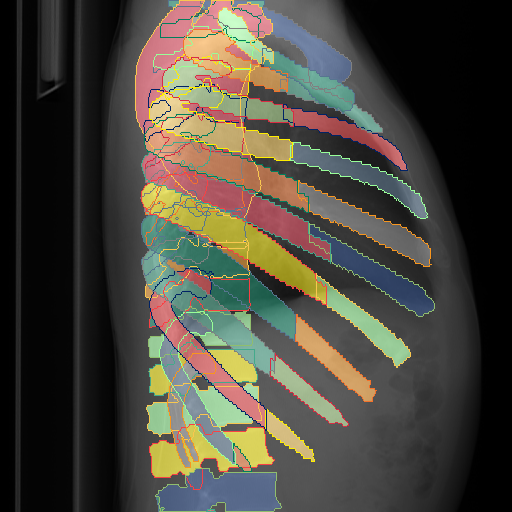}&
         \includegraphics[width=0.175\linewidth,height=0.175\linewidth]{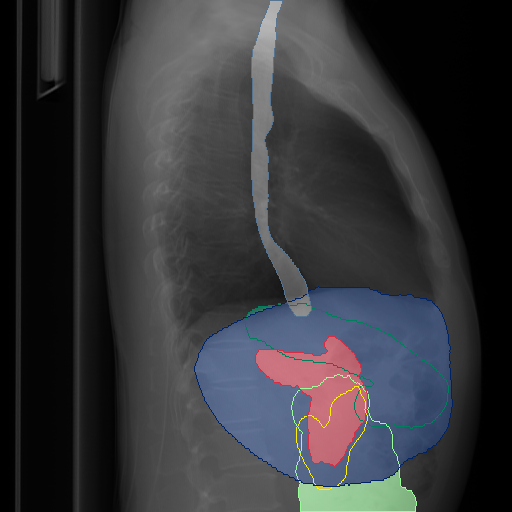}\\
         \rotatebox{90}{\phantom{0000}Covid19AR}& 
         \includegraphics[width=0.175\linewidth,height=0.175\linewidth]{figures/dataset_samples/COVID19AR_4_lateral.png}&
         \includegraphics[width=0.175\linewidth,height=0.175\linewidth]{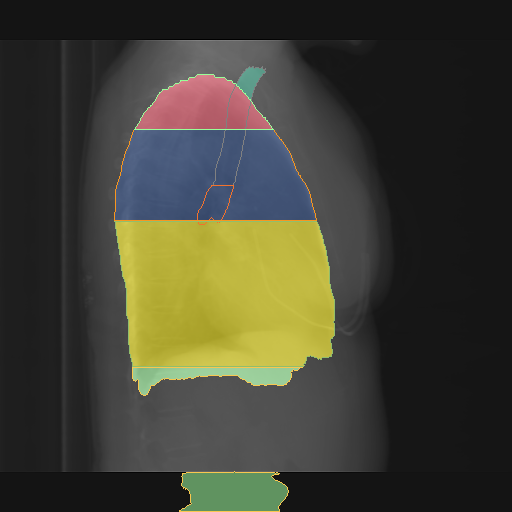}&
         \includegraphics[width=0.175\linewidth,height=0.175\linewidth]{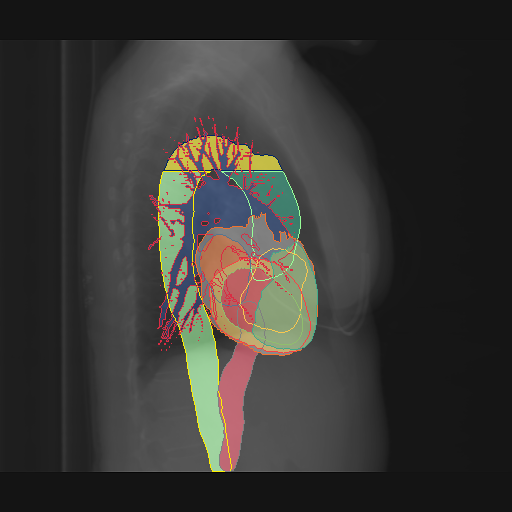}&
         \includegraphics[width=0.175\linewidth,height=0.175\linewidth]{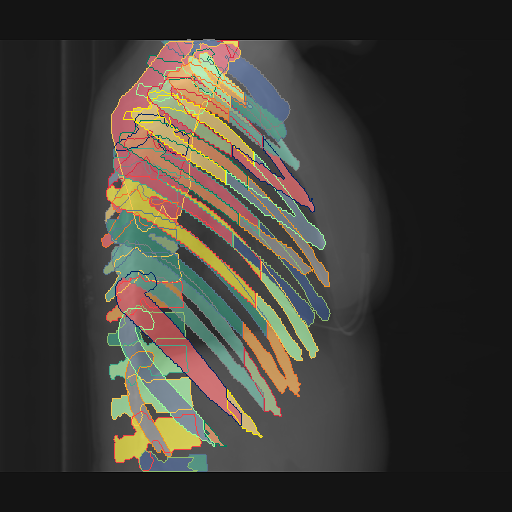}&
         \includegraphics[width=0.175\linewidth,height=0.175\linewidth]{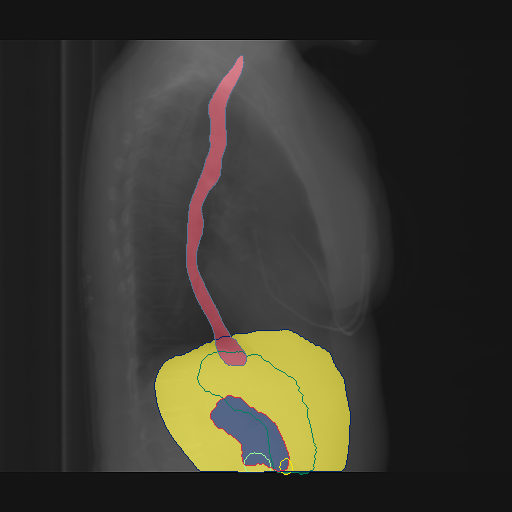}\\
         \rotatebox{90}{\phantom{0000}Covid1910}& 
         \includegraphics[width=0.175\linewidth,height=0.175\linewidth]{figures/dataset_samples/COVID1910_0355_lateral.png}&
         \includegraphics[width=0.175\linewidth,height=0.175\linewidth]{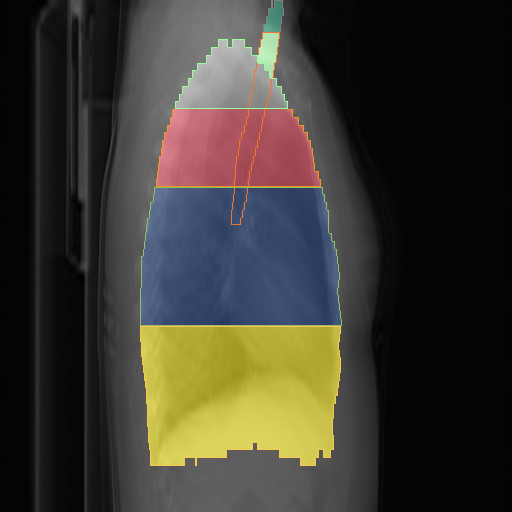}&
         \includegraphics[width=0.175\linewidth,height=0.175\linewidth]{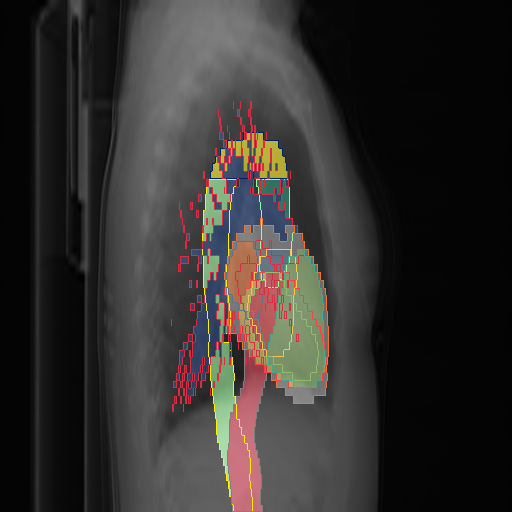}&
         \includegraphics[width=0.175\linewidth,height=0.175\linewidth]{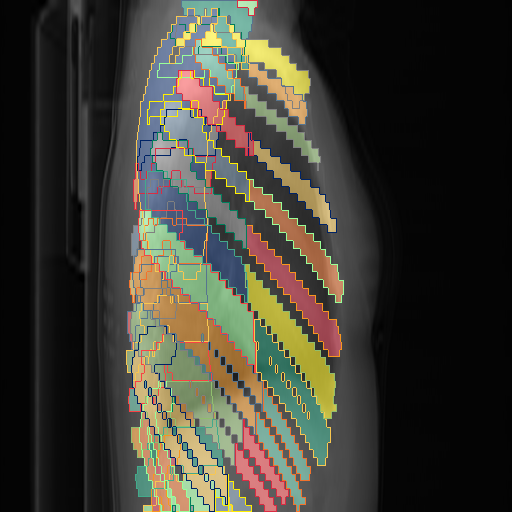}&
         \includegraphics[width=0.175\linewidth,height=0.175\linewidth]{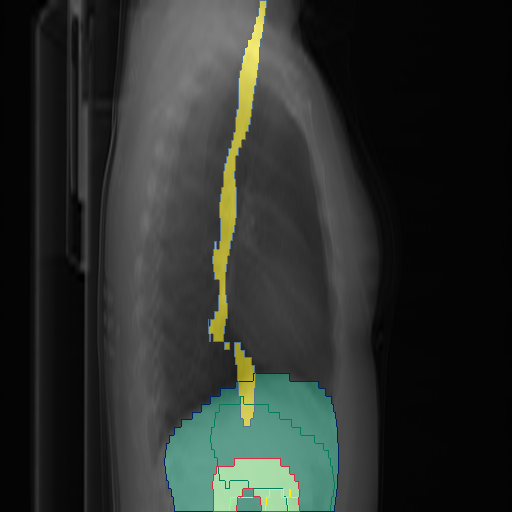}\\
         \rotatebox{90}{\phantom{0000}Covid1920}& 
         \includegraphics[width=0.175\linewidth,height=0.175\linewidth]{figures/dataset_samples/COVID1920_296_lateral.png}&
         \includegraphics[width=0.175\linewidth,height=0.175\linewidth]{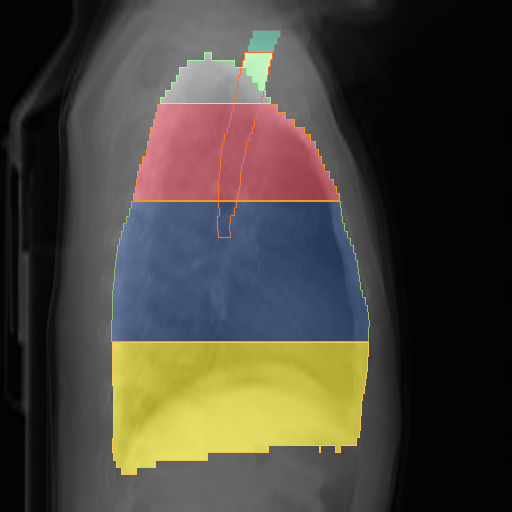}&
         \includegraphics[width=0.175\linewidth,height=0.175\linewidth]{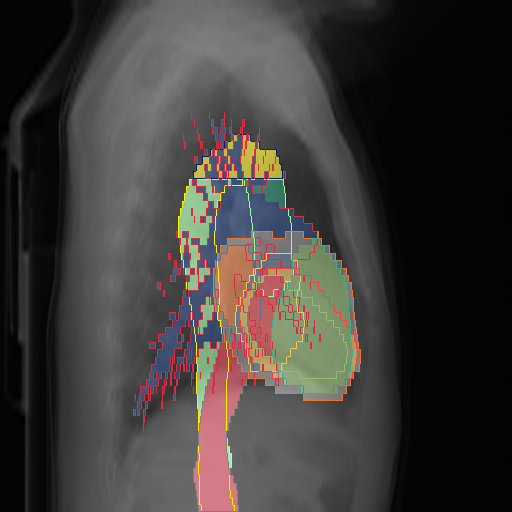}&
         \includegraphics[width=0.175\linewidth,height=0.175\linewidth]{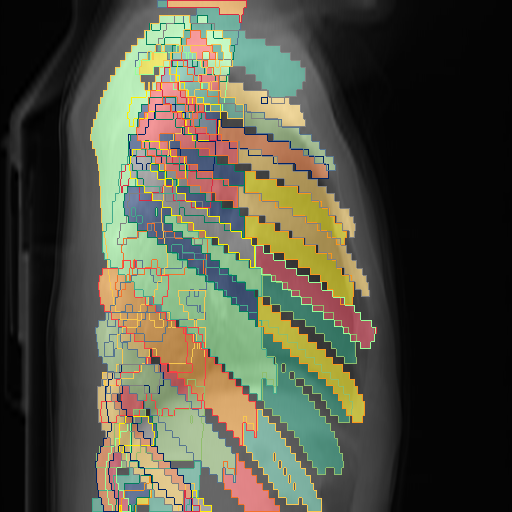}&
         \includegraphics[width=0.175\linewidth,height=0.175\linewidth]{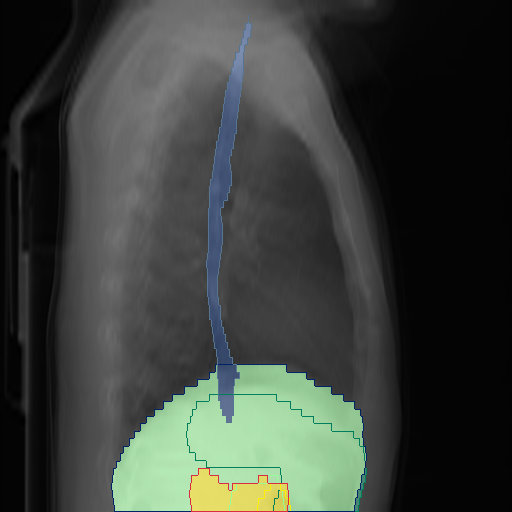}\\
         \rotatebox{90}{\phantom{00000}RSNA PE}& 
         \includegraphics[width=0.175\linewidth,height=0.175\linewidth]{figures/dataset_samples/RSNAPE_fc15f24c6ffb_aa4ad84d4ec6_lateral.png}&
         \includegraphics[width=0.175\linewidth,height=0.175\linewidth]{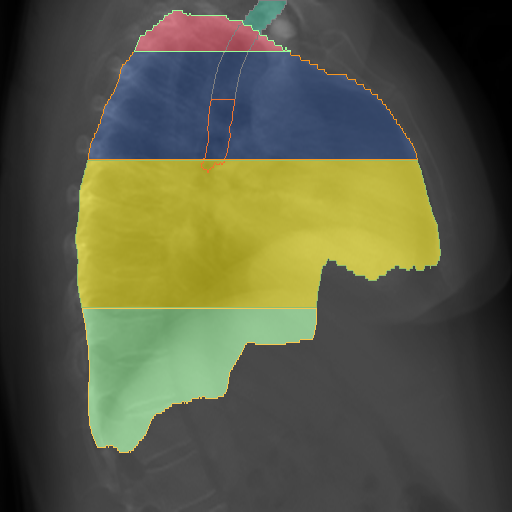}&
         \includegraphics[width=0.175\linewidth,height=0.175\linewidth]{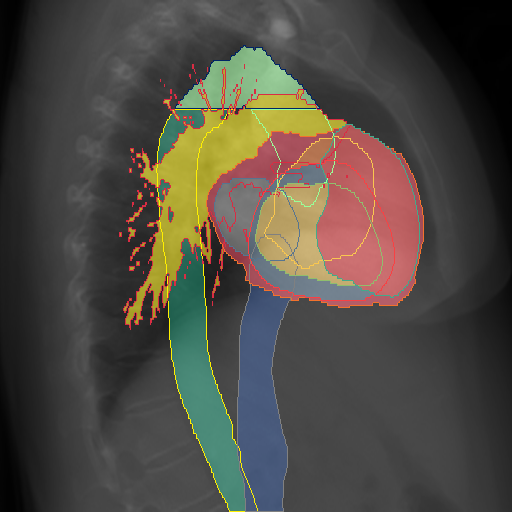}&
         \includegraphics[width=0.175\linewidth,height=0.175\linewidth]{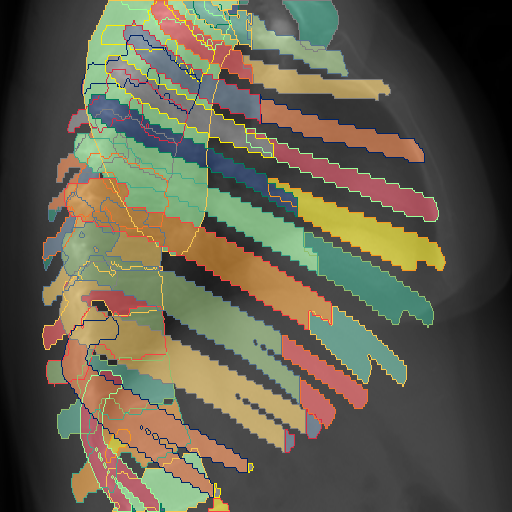}&
         \includegraphics[width=0.175\linewidth,height=0.175\linewidth]{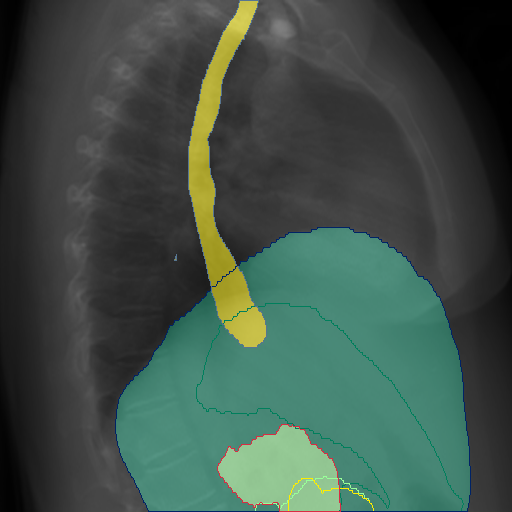}\\
         \rotatebox{90}{\phantom{0000000}LITS}& 
         \includegraphics[width=0.175\linewidth,height=0.175\linewidth]{figures/dataset_samples/1010b_lateral.png}&
         \includegraphics[width=0.175\linewidth,height=0.175\linewidth]{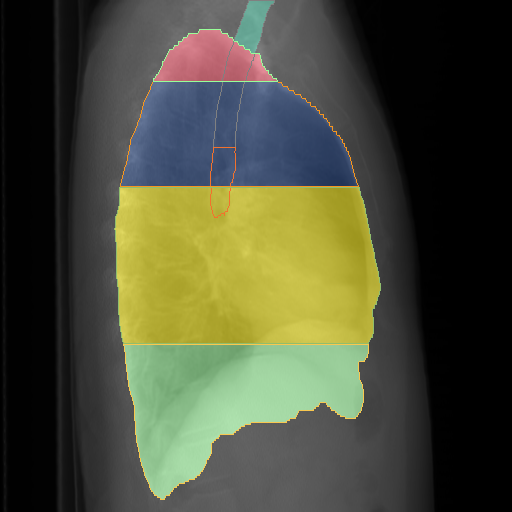}&
         \includegraphics[width=0.175\linewidth,height=0.175\linewidth]{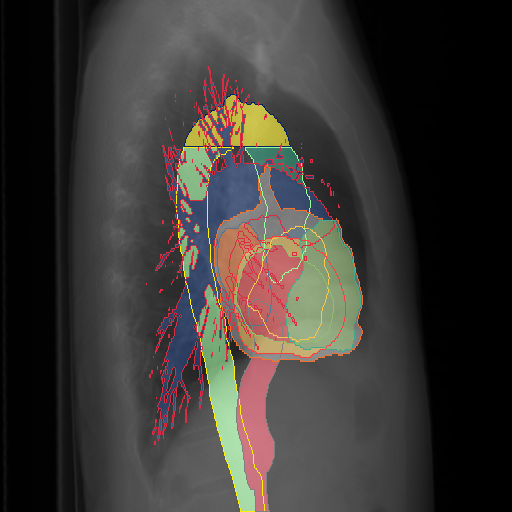}&
         \includegraphics[width=0.175\linewidth,height=0.175\linewidth]{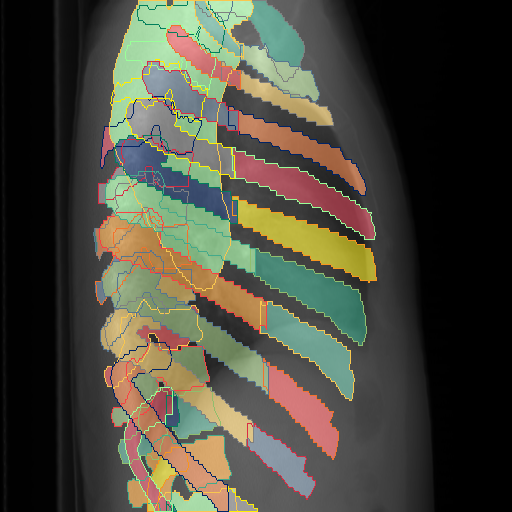}&
         \includegraphics[width=0.175\linewidth,height=0.175\linewidth]{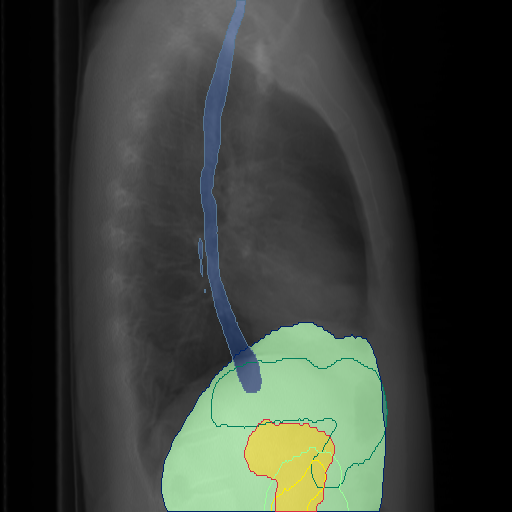}\\
         \bottomrule
    \end{tabular}
    \caption{Sample of lateral projected x-rays from the RibFrac, COVID-19-AR, COVID-19-1, COVID-19-2, RSNA PE, and LIDC-IDRI dataset. We show labels belonging to the categories \textit{Respiratory System}, \textit{Vascular System}, \textit{Bones}, and \textit{Abdomen}.}\label{fig:paxray_lateral}
\end{figure*}

Furthermore, when projecting the pseudo-labels to a frontal or lateral view, we merge classes to accommodate the characteristics of the view. For example, distinguishing between the left and right scapula is impossible in the 
lateral view, whereas this is trivial in the frontal view. On the other hand, the subsegments of the lower mediastinum is better distinguishable in the lateral view, while not in the frontal one. We display the final classes in Fig.~\ref{fig:labelstructure_extendedpaxray}.

\subsection*{Image Sources of PAX-Ray++}  
To generate the PAX-Ray++ dataset, we require CT volumes as imaging data. We choose large unlabeled CT datasets focusing on the thoracic region, which is displayed in the bottom part of Table~\ref{tab:ch5_datasets_overview}.

 The datasets include RSNA PE, RibFrac, LIDC-IDRI, COVID-19-AR, COVID-19-1, and COVID-19-2.

 \begin{enumerate}
     \item The RNSA Pulmonary Embolism dataset, a contrast CT dataset for identifying pulmonary embolisms, is the largest, with more than 7000 volumes with an average axial spacing of 1.25 mm and 237 slices.  
     \item  The RibFrac dataset for detecting rib fractures, which we used for the initial PAX-Ray, consists of 660 volumes with the smallest average spacing of 1.13 mm and 359 slices. 
     \item  The LIDC-IDRI dataset was collected to detect pulmonary nodules and consists of 1036 volumes with a spacing of 1.74 and, on average, 237 slices. 
     \item  The COVID-19-AR dataset has, on average, the most slices of 438 and 176 volumes. However, it has a relatively large coronal and sagittal spacing. 
     \item  The COVID-19-1 and -2 datasets contain 215 and 632 volumes but have the smallest amount of slices and the largest axial spacing. 
 \end{enumerate}

We infer the nnUNet ensemble trained on the aggregated pseudo-labels on Autopet on these datasets (see Fig.~\ref{fig:workflow}[a]). We apply the same postprocessing methods as done with the initial predictions on AutoPET and filter volumes with predictions with anatomical deviations, i.e., too few predicted ribs.
 We then project the image and label files to a frontal and lateral view and resize to a uniform size of $512\times512$ using nearest interpolation for masks and Lanczos~\cite{turkowski1990filters} for images.

We show the resulting masks in Fig.~\ref{fig:paxray_frontal} and Fig.~\ref{fig:paxray_lateral}. We see that the different characteristics of each CT dataset have a direct influence on the resulting projections. Thus, the resolution of each mask depends on the resolution of the original CT scan.

\subsection*{Properties of PAX-Ray++}
\begin{figure*}[t]
    \centering
    \includegraphics[width=\linewidth]{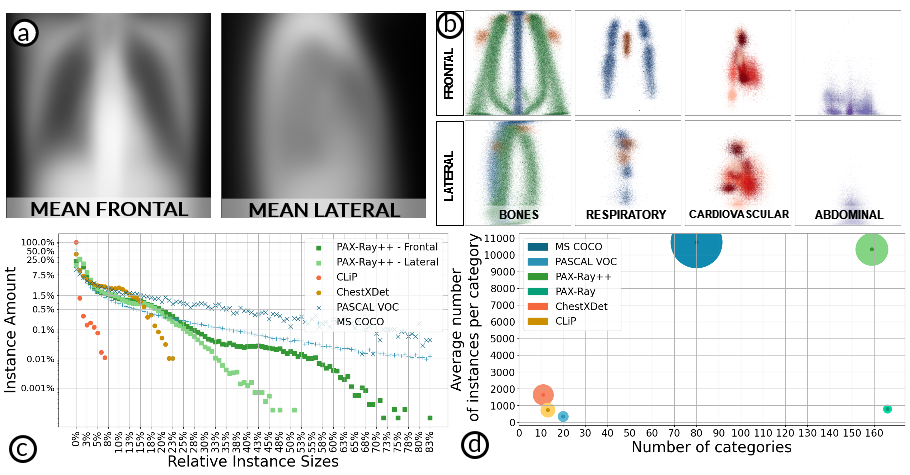}
    \caption{Illustration of dataset characteristics. (a) shows the mean images for the frontal and lateral view (b) shows centerpoint distributions of individual class annotations for different views and superclasses. (c) shows instance size comparisons between datasets. (d) Shows the average number of instances per class versus the number of classes for different datasets.}
    \label{fig:properties}
\end{figure*}

\noindent\textbf{Mean Images.} In Fig.~\ref{fig:properties}$\cir{a}$, we highlight the mean images of the lateral and frontal views in our dataset. All frontal and lateral images are homogenously oriented. Due to the choice of projection, the frontal view resembles an intermediate between the commonly used anterior-posterior (AP) and posterior-anterior (PA) views. The lateral view mean image all resemble left lateral CXR. Due to the difference in image acquisition, there are several structural differences between the projections and the real CXR. Since the CT portrays a patient lying down, several structures slightly press against the back. In contrast, the PA CXR shows the patient in an upright position pulling several organs toward the bottom.

\noindent\textbf{Class Center Distributions.} In \Cref{fig:properties}$\cir{b}$, we show the center-point distributions for several superclasses denoted with a corresponding color. We note that the effect of the provided crop in the original datasets leads to a difference in the position and scale of the thoracic anatomy after rescaling the projections to a uniform size. The frontal ribs in green illustrate this, as the center points of the individual anterior and posterior rib parts contain a noticeable separation at the bottom half of the image. We observe that the individual classes belonging to bones (1st column, vertebrae in blue, clavicles and scapulae in orange, ribs in green), abdomen (4th column, purple), and the respiratory (2nd column, lung sections in blue, trachea orange) and cardiovascular (3rd column, shades of red) systems appear in the expected positions for a standard CXR in both frontal and lateral views.
\newline 

 \noindent\textbf{Comparison of Datasets.} We compare our PAX-Ray++ dataset against other datasets within and outside the CXR domain in \Cref{fig:workflow}$\cir{c}$/$\cir{d}$. This includes ChestXDet~\cite{liu2020chestxdet10}, a dataset with mask annotations for lung diseases, CLiP~\cite{tang2021clip}, a dataset with annotations for catheters and lines, our PAX-Ray, as well as the popular natural image datasets PASCAL VOC~\cite{pascal-voc-2007} and MSCOCO~\cite{lin2014microsoft}. PASCAL VOC and MSCOCO are benchmark datasets that support the development of a wide range of computer vision algorithms. We consider them to visualize the existing gap between the natural image and the CXR domain.

\noindent\textbf{Instance Size Distributions.} In Fig.~\ref{fig:properties}$\cir{c}$, we plot the size of each instance annotation relative to the image size. 
The natural image datasets show more instances that take up more prominent parts of the image, with at least 75\% and 50\% of annotations taking up more than 1\% of the image for PASCAL VOC and MS COCO respectively. On the other hand, CXR datasets mainly consist of smaller annotations. 
We see that nearly all annotations of CLiP are at most 1\% of the image size and at most 0.01\% of annotations taking up 7\%. ChestXDet is more balanced. However, less than 0.1\% of annotations take up more than 20\% of the image. Related to the domain, PAX-Ray++ also consists primarly of more minor instances. However, it also shows instances of up to 70\% of the image.

\noindent\textbf{Instance Annotations.} In Fig.~\ref{fig:properties}$\cir{d}$, we compare the number of instances per category against the number of categories. We see that within the domain, PAX-Ray++ provides significantly more different annotations and, in general, provides more annotations than other datasets within the domain, even comparable to natural image datasets like MS COCO while providing more classes. We note that images in natural image domains occur in vastly different contexts, while CXR images are roughly homogeneous due to the image acquisition process.

\subsection*{Chest X-Ray Segmentation}
For our segmentation models, we chose the UNet~\cite{unet} with a ResNet-50~\cite{he2016deep} and ViT-B~\cite{dosovitskiy2020image} backbone. We used pre-trainings on ImageNet~\cite{imagenet} as we did not see a difference in performance to other pre-training methods, i.e., MAE~\cite{he2022masked} on CXRs. However, as shown in previous work~\cite{Seibold_2022_BMVC}, pre-training had significantly improved results compared to randomly initialized weights.
We train the network with binary cross-entropy and employ an additional
binary dice loss which we optimize using AdamW~\cite{loshchilov2017decoupled} with a learning rate of 0.001 for
110 epochs decaying by a factor of 10 at \{60, 90, 100\} epochs. For our base augmentation, we used random resize-and-cropping of range [0.8, 1.2] as base augmentation for an image size of 512. We utilize RandAugment~\cite{cubuk2020randaugment} with their proposed augmentations. We set $N=9$ and randomized magnitude for heavy augmentation.

\section*{Examples of Limitations}
\begin{figure}
    \centering
    \begin{tabular}{ccc}
         \includegraphics[width=0.3\linewidth]{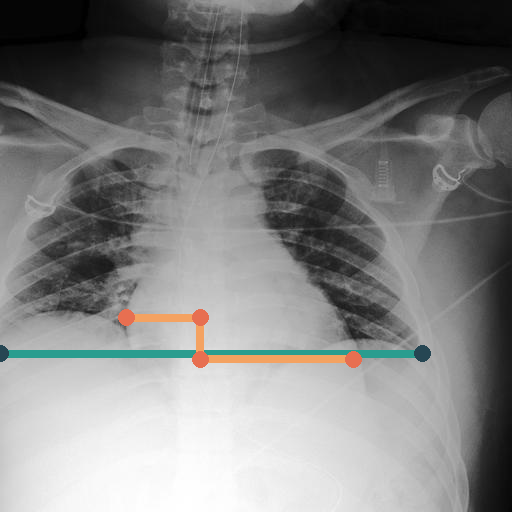}&\includegraphics[width=0.3\linewidth]{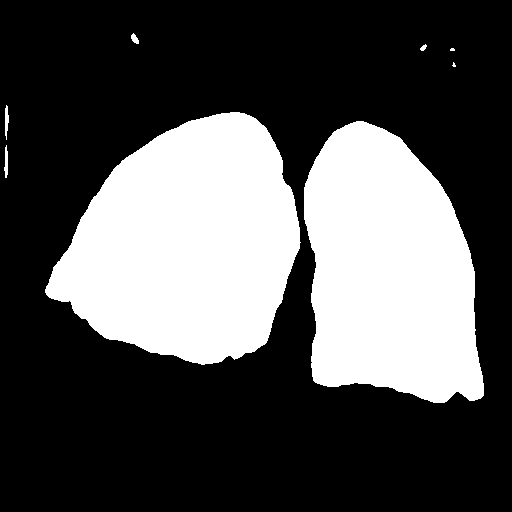}&\includegraphics[width=0.3\linewidth]{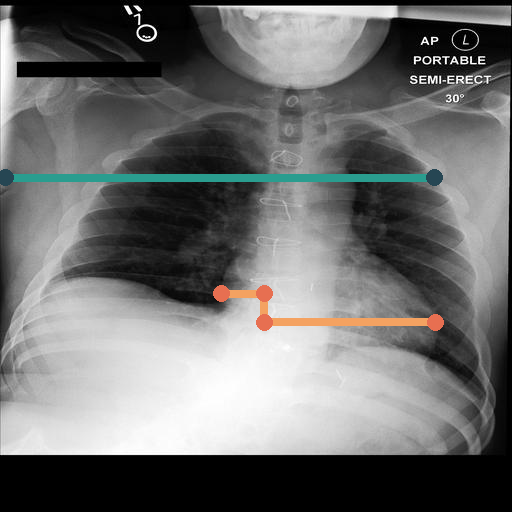}  \\
          \includegraphics[width=0.3\linewidth]{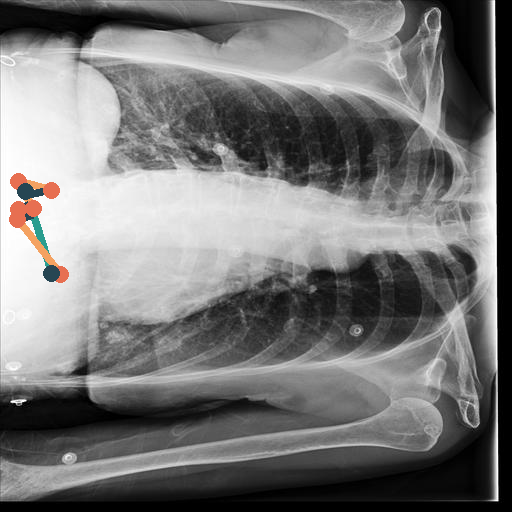}&\includegraphics[width=0.3\linewidth]{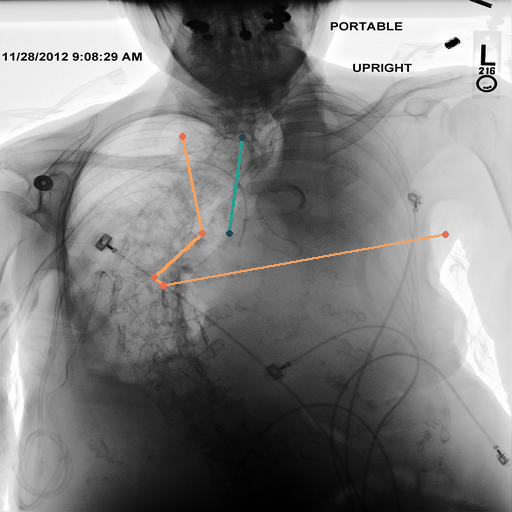} &\includegraphics[width=0.3\linewidth]{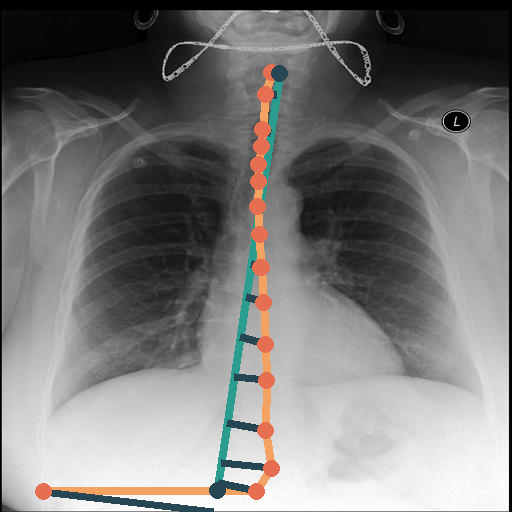}
    \end{tabular}
    \caption{We show limitations our method for understandable feature extraction via CXAS. The top row shows failure cases for the Cardio-Thoracic Ratio. 
    Errors can be generated through unfortunate image acquisition cropping the image (left column) or speckles within the prediction mask leading to a missleading extraction of the internal diameter. 
    The bottom row shows failure cases for the Spine-Center Distance. Unseen rotations, inverted imaging and indistinguishable abdominal areas lead to segmentation errors of the  entire spine or individual vertebrae. }
    \label{fig:limitations}
\end{figure}

We show limitations of CXAS for understandable feature extraction in Fig.\ref{fig:limitations}. The top row shows failure cases for the Cardio-Thoracic Ratio. Errors can be generated through unfortunate image acquisition cropping the image (left column) or speckles within the prediction mask leading to a missleading extraction of the internal diameter. The bottom row shows failure cases for the Spine-Center Distance. Unseen rotations, inverted imaging and ndistinguishable abdominal areas lead to
segmentation errors of the entire spine or individual vertebrae.

\section*{Compliance with ethical standards:}
The used datasets are publicly available. Ethical approval was not required as
confirmed by the license attached with the open access data.

\section*{Acknowledgements}
The present contribution is supported by the Helmholtz Association under the joint research school “HIDSS4Health – Helmholtz Information and Data Science School for Health” and by the Helmholtz Association Initiative and Networking Fund on the HAICORE@KIT partition.
\section*{Data Availability}
The PAX-Ray++ dataset is available \href{https://drive.google.com/drive/folders/1AEJAaPTxVMx9iofY4J4f2x5gpJqE61I2?usp=sharing}{here}.
\section*{Code Availability}
Our code is available \href{https://github.com/ConstantinSeibold/ChestXRayAnatomySegmentation}{here}. We make our CXAS model available \href{https://drive.google.com/drive/folders/1JLmbkN-n50giwk60e25oqi_JAG6WfAOQ?usp=sharing}{here} and usable via \texttt{pip install cxas}.
\section*{Author Contributions}
C.S. designed the method, trained 2D/3D models, merged annotations, generated the 2D and 3D dataset, conducted and interpreted experiments and wrote the article. A.J. trained 3D models and merged annotations. M.A.F. and M.-S.K. gave clinical feedback. M.A.F. and S.R. gave advice and edited the article. J.K. and R.S. supervised the research.


\end{document}